\documentclass[twocolumn,trackchanges]{aastex63}

\usepackage{amsmath}
\usepackage{graphicx}
\usepackage{soul}

\newcommand{\ramses}{\textsc{ramses}}
\newcommand{\ramsesrt}{\textsc{ramses-rt}}
\newcommand{\makedisk}{\textsc{makedisk}}

\definecolor{red}{RGB}{190,0,0}

\definecolor{pink}{RGB}{255,51,153}

\definecolor{green}{RGB}{0,153,0}

\newcommand{\msun}{\mbox{$\rm M_{\odot}$}}
\newcommand{\msunyr}{\mbox{$\rm M_{\odot}\,{\rm yr^{-1}}$}}
\newcommand{\kms}{\mbox{${\rm km\,s^{-1}}$}}

\newcommand{\cmq}{\mbox{$\,{\rm cm^{-3}}$}}
\newcommand{\nH}{\mbox{$n_{\rm H}$}}

\received{\today}
\revised{\today}

\shorttitle{Ram Pressure Stripping}
\shortauthors{Jaehyun Lee et al.}

\begin{document}
\title{Dual Effects of Ram Pressure on Star Formation in Multi-phase Disk Galaxies with Strong Stellar Feedback}

\author[0000-0002-6810-1778]{Jaehyun Lee}
\affiliation{Korea Institute for Advanced Study, 85, Hoegi-ro, Dongdaemun-gu, Seoul 02455, Republic of Korea}

\author[0000-0002-3950-3997]{Taysun Kimm}
\affiliation{Department of Astronomy, Yonsei University, 50 Yonsei-ro, Seodaemun-gu, Seoul 03722, Republic of Korea}

\author{Harley Katz}
\affiliation{Astrophysics, University of Oxford, Denys Wilkinson Building, Keble Road, Oxford OX1 3RH, UK}

\author[0000-0002-7534-8314]{Joakim Rosdahl}
\affiliation{Univ Lyon, Univ Lyon1, Ens de Lyon, CNRS, Centre de Recherche Astrophysique de Lyon UMR5574, F-69230 Saint-Genis-Laval, France}

\author[0000-0002-8140-0422]{Julien Devriendt}
\affiliation{Astrophysics, University of Oxford, Denys Wilkinson Building, Keble Road, Oxford OX1 3RH, UK}

\author{Adrianne Slyz}
\affiliation{Astrophysics, University of Oxford, Denys Wilkinson Building, Keble Road, Oxford OX1 3RH, UK}

\email{syncphy@gmail.com,tkimm@yonsei.ac.kr}

\begin{abstract}
We investigate the impact of ram pressure stripping due to the intracluster medium (ICM) on star-forming disk galaxies with a multi-phase interstellar medium (ISM) maintained by strong stellar feedback. We carry out radiation-hydrodynamics simulations of an isolated disk galaxy embedded in a $10^{11}\,\msun$ dark matter halo with various ICM winds mimicking the cluster outskirts (moderate) and the central environment (strong). We find that both star formation quenching and triggering occur in ram pressure-stripped galaxies, depending on the strength of the winds. HI and H$_2$ in the outer galactic disk are significantly stripped in the presence of the moderate winds, whereas turbulent pressure provides support against ram pressure in the central region where star formation is active. Moderate ICM winds facilitate gas collapsing,  increasing the total star formation rates by $\sim 40\%$ when the wind is oriented face-on or $\sim80\%$ when it is edge-on. In contrast, strong winds rapidly blow away neutral and molecular hydrogen gas from the galaxy, suppressing the star formation by a factor of two within $\sim 200 \, {\rm Myr}$. Dense gas clumps with $N_{\rm H} \ga 10\, {\rm M_{\odot}\, pc^{-2}}$ are easily identified in extraplanar regions, but no significant young stellar populations are found in such clumps. In our attempts to enhance radiative cooling by adopting a colder ICM of $T=10^6\,{\rm K}$, only a few additional stars are formed in the tail region, even if the amount of newly cooled gas increases by an order of magnitude.
\end{abstract}

\section{Introduction}

Cluster galaxies in the local Universe are redder, gas-poorer, and tend to be more elliptical than those in the field \citep[e.g.,][]{butcher78,dressler80,cortese07,park09a,peng10,mahajan12,johnston14,fasano15}. Strong tidal fields induced by deep the gravitational potential or frequent interactions transform galaxy morphology and facilitate gas loss from the galaxies \citep[e.g.,][]{icke85,byrd90,moore96,barnes96}. The hot intracluster medium (ICM) can gradually evaporate the interstellar medium (ISM) by thermal conduction \citep{cowie77} and exert ram pressure, which strips the circumgalactic medium (CGM) and ISM away \citep{gunn72,davies73}. All of these mechanisms are more efficient in clusters than in more isolated environments, suggesting that environmental effects disturb stellar kinematics and  quench star formation.

Numerous attempts have been made to uncover direct evidence for these environmental effects on galaxies. Radio observations based on the hydrogen 21-cm line show that galaxies residing in denser environments are more deficient in neutral hydrogen (HI) \citep{davies73,giovanelli85,giraud86,warmels88b,cayatte90,scodeggio93,solanes01}. High-resolution 21-cm imaging with the Very Large Array revealed extra-planar HI distributions and long HI tails behind galaxies in a cluster, which is a clear signature of ram pressure stripping \citep[e.g.,][]{bravo-alfaro00,kenney04,chung09,scott10}.  In addition to this, using Multi-Unit Spectroscopic Explorer observations of a spiral galaxy falling into the center of a cluster, \citet{fumagalli14} observed stripped features in H$\alpha$ emission without any sign of disturbed velocity fields of stellar components, which would otherwise be the evidence of the effects of tidal fields. In fact, 
the presence of ionized tails is often observed \citep[e.g.][]{gavazzi01,boselli14b,yagi17}, demonstrating the pivotal role of ram pressure stripping in cluster environments. Although very rare, massive molecular tails, which are speculated to form {\it in-situ}, have also been discovered behind a couple of ram pressure stripped galaxies \citep{jachym17,jachym19}. These tails are HI deficient while bright in H$\alpha$ and X-ray, possibly indicating that they may be heated by the hot ICM. 

Although disrupted features are often observed in HI or H$\alpha$ in the disks of ram pressure stripped galaxies, this seems to be hardly the case for molecular disks. As mentioned above, molecular tails are detected in a few galaxies to date, but their origin remains unclear. Some authors have claimed that galaxies in the cluster center have a smaller amount of molecular hydrogen \citep{rengarajan92}, but a significant number of studies have argued that galaxies still maintain their molecular gas even after falling into the cluster center \citep{stark86,kenney89,crowl05,vollmer08,abramson14,kenney15,lee17}. Based on Submillimeter Array observations of the three Virgo spiral galaxies, which are known to have ram pressure-stripped HI disks \citep{chung09}, \citet{lee17} showed that the spirals have a CO distribution (indicating the presence of H$_2$), which is more concentrated than that of HI, with morphological peculiarities, but without any clear evidence of molecular gas stripping. Interestingly, \citet{lee17} also found that the distribution of far-ultraviolet emission, which traces star formation over the past $\sim 100$ Myr \citep{kennicutt98}, is significantly more extended than the CO and H$\alpha$ emission, possibly hinting at shrinking molecular disks due to ram pressure \citep[see also][]{koopmann04b,cortese06}.

However, although ram pressure from the ICM may eventually transform cluster satellites into red and dead galaxies by blowing HI away \citep{koopmann04a,koopmann04b}, the suppression of star formation may not always occur immediately, especially in the early stage of ram pressure stripping. In some cluster satellites, dubbed as `jellyfish galaxies', star-forming knots are found in their wakes \citep[e.g.][]{sun07,owers12,fumagalli14,ebeling14,rawle14,poggianti16,sheen17}, suggesting that star formation may be triggered by ram pressure, facilitating the formation of intracluster light \citep[e.g.,][]{mihos17,ko18}. 

Unravelling the complex effects of ram pressure stripping on star formation in cluster environments has been a key goal of theoretical studies. Ever since \citet{gunn72} developed an analytic model for the truncation of a gaseous disk due to ram pressure, several authors have investigated the stripping process and confirmed the analytic predictions using Lagrangian and Eulerian hydrodynamics simulations \citep[e.g.,][]{abadi99,schulz01,quilis00,roediger05}.
Studies have also shown that gaseous disks are stripped more easily when galaxies orbit more rapidly, encounter a denser ICM, or have spin axes better aligned with the direction of the motion \citep{schulz01, vollmer01,roediger06a,vollmer06,roediger07,roediger08,jachym09,jung18}. \citet{shin13} further demonstrated that longer and smoother tails develop behind galaxies when the galactic gas is more turbulent. Recent studies based on cosmological simulations argue that more than $\sim 30\%$ of satellite galaxies in massive groups and clusters exhibit ram-pressure-stripped tails, and that this fraction should increase with lower stellar mass \citep[e.g.][]{yun19}. However, no consensus has been reached yet with regard to star formation activities of the ram-pressure-stripped galaxies.
Some claim that ram pressure enhances star formation by compressing gaseous disks in the early phase of the ICM-ISM interaction \citep{schulz01,vollmer01,bekki03,kronberger08,kapferer08, kapferer09, steinhauser12} or by triggering star formation in the ram pressure stripped tails \citep{bekki03,kronberger08,kapferer08,kapferer09, steinhauser12,tonnesen12}. In contrast, \citet{tonnesen12} showed that the star formation rates diminish in the galactic plane right after the ICM wind hits the galaxy. These results indicate that galactic star formation responds to ram pressure stripping in various ways, and therefore a further systematic study is required to probe the impact of ram pressure stripping.

Generally, the low-density ISM is stripped relatively easily, while clouds located at the central regions of galaxies are barely affected by cluster-like environments \citep[e.g.][]{quilis00,tonnesen09}, which indicates that the multi-phase nature of the ISM may be critical for the accurate understanding of the effects of ram pressure \citep[c.f.,][]{tonnesen10}. Furthermore, a number of observations have reported that stripped tails are also multi-phase \citep{kenney04,oosterloo05,chung07,chung09,jachym14,verdugo15,jachym17,moretti18}, and they seem to originate from stripped clouds and clouds formed {\it in-situ} \citep[e.g.,][]{jachym19}. Simulations with stellar feedback can easily develop a three-phase ISM \citep{mckee07}, but, to the best of our knowledge, no ram pressure stripping simulation has been performed at high resolution and with stellar feedback that is strong enough to regulate star formation as is required by observations. In addition, simulations integrating the complex chain of cooling, non-equilibrium chemistry, star formation, and feedback are lacking. Therefore, this study aims to examine the impact of ram pressure stripping on star formation activity and a multi-phase disk using a radiation-hydrodynamic code that can trace the evolution of the ISM with strong feedback, photo-chemistry, and thermo-turbulent star formation models \citep[][]{kimm15,katz17,kimm17}.

This paper is organized as follows. Section 2 describes the radiation-hydrodynamics method and initial conditions of the simulations. In Section 3, we present the evolution of gaseous disk, the properties of the ram pressure stripped tails, and star formation activity. Section 4 discusses the dual effects of ram pressure on star formation, role of turbulent pressure on the stripping process, the lack of extraplanar star formation, and potential shortcomings of our approaches. Finally, we summarize our major findings in Section 5.

\section{Simulation}
 To investigate the impact of ram pressure stripping on star forming galaxies, we carry out a set of idealized radiation-hydrodynamics simulations. In this section, we describe physical processes included in our simulations and how their initial conditions are set up.

\subsection{Code}

We use \ramsesrt\ \citep{rosdahl13,rosdahl15a}, a radiation-hydrodynamics version of the adaptive mesh refinement code \ramses\ \citep{teyssier02}, to investigate the impact of ram pressure stripping on star-forming galaxies. The Euler equations are solved with a Courant number of 0.8 using the HLLC solver \citep{toro94}, and the Poisson equation is solved using the Particle-Mesh method \citep{guillet11}. The publicly available version of \ramsesrt\ solves the non-equilibrium chemistry and cooling of the six chemical species, HI, HII, HeI, HeII, HeIII, and e$^{-}$ \citep{rosdahl13,rosdahl15a}. We use a modified photo-chemistry model for the primordial species to include the formation and destruction of molecular hydrogen, as detailed in Section~\ref{sec:h2}. We use a reduced speed of light ($10^{-3}\,c$) to reduce the computational cost.  We use the default Cloudy \citep{ferland98} cooling model in \ramsesrt\ (\texttt{cc07}) for  atomic metal cooling at $T\ga 10^4\,{\rm K}$, and \citet{rosen95} for the fine-structure line cooling at  $T\la 10^4\,{\rm K}$. We also include the cooling due to molecular hydrogen using the cooling rates from \citet{hollenbach79,halle13}, as described in \citet{kimm17}.  Heating due to the uniform background UV radiation is included with a self-shielding approximation \citep{rosdahl13}, assuming the \citet{haardt12} spectrum at $z=0$. We do not impose any effective equation of state or pressure floor in this work. 

\subsubsection{Star formation}
Star formation rates are computed based on a Schmidt law \citep{schmidt59}, as
\begin{equation}
\label{eqn:schmidt}
\frac{{\rm d}\rho_{\rm star}}{{\rm d}t}=\epsilon_{\rm ff}\frac{\rho_{\rm gas}}{t_{\rm ff}},
\end{equation}
where $\rho_{\rm gas}$ is the gas density, $t_{\rm ff}=\sqrt{(3\pi/32G\rho_{\rm gas})}$ is the free fall time, $G$ is the gravitational constant, and $\epsilon_{\rm ff}$ is the star formation efficiency per free-fall time. Motivated by recent studies that suggest $\epsilon_{\rm ff}$ is not a constant but depends on the physical properties of the ISM \citep{padoan11,federrath12}, our star formation model calculates $\epsilon_{\rm ff}$ based on a local thermo-turbulent condition \citep{kimm17,trebitsch17}, as
\begin{equation}
\label{eqn:eff}
\epsilon_{\rm ff}=\frac{\epsilon_{\rm ecc}}{2\phi_{\rm t}} \exp \bigg( \frac{3}{8} \sigma^2_s \bigg) \bigg[ 1+ {\rm erf} \bigg( \frac{\sigma^2_s-s_{\rm crit}}{\sqrt{2\sigma^2_s}}\bigg) \bigg],
\end{equation}
where $\epsilon_{\rm ecc}\approx0.5$ is the fraction of gas that can be accreted on to stars without being affected by proto-stellar jets and outflows, $1/\phi_{\rm t}\approx0.57$ is a numerical factor that accounts for the multi-free-fall time in the clouds, $\sigma^2_s= \ln (1+b^2\mathcal{M}^2)$ is the standard deviation of the logarithmic density contrast ($s\equiv\ln(\rho/\rho_0)$) where $b$ is a parameter that depends on the mode of turbulence driving, which we select as $b=0.4$ for the mixture of solenoidal and compressive turbulence, $\mathcal{M}$ is the sonic Mach number, $\rho_0$ is the mean density of the cloud, and  the critical density above which gas may start to collapse ($s_{\rm crit}$) is approximated \citep{federrath12}, as
\begin{equation}
\label{eqn:scrit}
s_{\rm crit}=\ln(0.067\theta^{-2}\alpha_{\rm vir}\mathcal{M}^2),
\end{equation}
where $\theta=0.33$ is a numerical factor for the uncertainty in the post-shock thickness with respect to the cloud size, $\alpha_{\rm vir}=2E_{\rm kin}/|E_{\rm grav}|$ is the virial parameter of the cloud. We estimate the local Mach number from the velocity dispersion between neighbouring grid cells defined by $\sigma_{\rm gas}^2={\rm Tr} \left( \vec{\nabla} v^{\rm T} \vec{\nabla} v \right) \Delta x^2$, where $v$ is the divergence and rotation-free local velocity field.  Note that the model is designed to form stars preferentially in regions where the sum of thermal and turbulent pressure is not strong enough to counter-balance the gravitational collapse of gas clouds. The resulting typical $\epsilon_{\rm ff}$ used to form a star particle in our simulations is $\sim0.2$. The typical density at which star particles are formed in our simulations is $\nH\sim800\,\cmq$, which is significantly higher than the threshold density that we set for star formation ($n_{\rm H,thres}=100\,\cmq$). Once the efficiency is determined, we use a Poisson distribution to randomly sample the mass of a star particle with the minimum mass of $M_{\rm star,min}=914\,\msun$, following \citet{rasera06}.

\subsubsection{Stellar feedback}

We include three different stellar feedback mechanisms that can regulate star formation in galaxies, i.e. photoionization, radiation pressure exerted by photons in the wavelength ranging from ultraviolet to infrared, and Type II supernova (SN) explosions.

Specifically, Lyman continuum photons produced by young massive stars can ionize hydrogen and heat up the gas to $T\sim 10^4\,{\rm K}$. This forms over-pressurized HII bubbles that can expand into the surrounding medium \citep[e.g.,][]{krumholz07,walch12,dale14}. During this process, absorbed ionizing photons transfer their momentum to the neutral ISM or dust \citep[e.g.][]{haehnelt95} and are assumed to be re-emitted in the IR range ($E< 1\,{\rm eV}$). We include the momentum transfer by computing the absorption due to neutral hydrogen ($E>13.6\,{\rm eV}$), neutral helium ($E>24.59\,{\rm eV}$), singly ionized helium ($E>54.42\,{\rm eV}$), and dust ($E>1\,{\rm eV}$) \citep{kimm17}. Multiple scattering of IR photons can also exert non-thermal pressure if they are efficiently trapped by optically thick dusty gas \citep{rosdahl15a}.  Photoelectric heating by UV photons with $5.6<E<13.6\,{\rm eV}$ is modeled following \citet{kimm17}. The photon production rates from stars are calculated based on the Binary Population and Spectral Synthesis model \citep[][v2.0]{stanway16} assuming a Kroupa initial mass function \citep[IMF;][]{kroupa01}  with a lower and upper mass cut-off of $0.1$ and $100\,\msun$, respectively.

The explosions of massive ($M\geq 8\msun$) stars are modelled as  a single event per star particle, based on the mechanical supernova (SN) feedback scheme developed by \citet{kimm14}. We adopt the final radial momentum of $p_{\rm rad}=4.5\times10^5\,\msun\,\kms$ per explosion if the host cell of SN is smaller than the Stromgren sphere \citep{geen15}, while a radial momentum of $p_{\rm rad}=2.5\times10^5\,\msun\,\kms$ \citep{thornton98} is used if the Stromgren sphere is well resolved, following the method described in \citet{kimm17}. If multiple SNe explode in the same time in one cell, we compute the final radial momentum with dependencies of metallicity ($Z'^{-2/17}$) and the number of SNe ($E_{51}^{16/17}$), where $Z'\equiv \min\left(Z/Z_{\odot},0.01\right)$ \citep{kimm14}. We assume that the SNe explode 5 Myr after the birth of a star particle. The frequency of SN explosions ($0.05\,\msun^{-1}$) is boosted by a factor of 5 in order to suppress over-cooling \citep[e.g.,][]{rosdahl17,rosdahl18}. Note that similar boost factor is employed in cosmological simulations to reproduce the luminosity function at the epoch of reionization \citep{rosdahl18} or the stellar-to-halo mass ratio of the Milky Way-like galaxies \citep{li18}. Finally, we assume that no metals are synthesized by SNe (i.e. zero stellar yield) in order to disentangle the contribution from the ICM wind to the formation of ram pressure stripped tails based on metallicity.

\subsubsection{Formation and destruction of H$_2$}
\label{sec:h2}

In the early Universe where little dust is present, the dominant route to produce H$_2$ is via H$^-$ \citep[e.g.,][]{abel97}.  However, in metal-rich environments, like the one modelled in this work, H$_2$ can be efficiently formed on dust grain surfaces \citep[e.g.][]{gould63}. We include the two formation channels, based on the prescription of \citet{glover10} and \citet{baczynski15}, as
\begin{align}
\label{eqn:h2rate}
\frac{{\rm d}x_{\rm H_2}}{{\rm d}t} & =\left(R_{\rm d} + R_{\rm p}\right)n_{\rm HI}-(k_{\rm LW}+k_{\rm UV})x_{\rm H_2} \nonumber \\
& -(C_{\rm coll, HI}+C_{\rm coll, H_2}+C_{\rm coll, HeI}+C_{{\rm coll}, e})x_{\rm H_2}
\end{align}
where  $x_{\rm H_2}\equiv n_{\rm H_2}/n_{\rm H}$ is the number fraction of molecular hydrogen, $n_{\rm HI}$ is the number density of HI, and $R_{\rm d}$ and $R_{\rm p}$ are the formation rate coefficients of H$_2$, in units of ${\rm cm^3\,s^{-1}}$, via  dust and H$^-$, respectively. 
$R_{\rm d}$ is computed as \citep{gnedin09}
\begin{equation}
\label{eqn:h2form}
R_{\rm d}=3.5 \, \times \, 10^{-17}~{\rm cm^3\,s^{-1}} \,  Z_{\rm gas} \, C_{\rho},
\end{equation}
where $Z_{\rm gas}$ is the gas metallicity, and $C_{\rho}$ is the gas clumping factor, and we adopt $C_{\rho}=3$ to reproduce the transition between neutral and molecular hydrogen (see Appendix A). We also take into account the formation via the H$^-$ process \citep[$R_p$, see Equations 10--11 in][]{katz17}, but the reaction is known to be very inefficient and does not increase the fraction of H$_2$ to more than $\sim 10^{-3}$ \citep[e.g.][]{bromm02}. 

Hydrogen molecules are photo-dissociated by Lyman-Werner radiation ($11.2~{\rm eV}\le E\le 13.6 ~{\rm eV}$)  emitted from stellar particles or destroyed by photons with $E\geq15.2~{\rm eV}$  through dissociative recombination. The photo-dissociation and photo-ionization rates due to Lyman-Werner and UV radiation, $k_{\rm LW} (=\sigma_{\rm LW} \mathcal{F}_{\rm LW})$ and $k_{\rm UV}(=\sigma_{\rm UV} \mathcal{F}_{\rm UV})$, are calculated using the local photon flux ($\mathcal{F}$) and associated cross-sections ($\sigma$) taken from \citet{glover08}.
H$_2$ molecules can also be destroyed by the collisions with other species (HI, H$_2$, HeI, and e$^-$). This is included by adopting the  temperature and density-dependent collisional dissociation rates,  $C_{{\rm coll}, i}$, from \citet{glover08}. Interested readers are referred to \citet{katz17} for details.

\subsection{Initial conditions}

The initial conditions of an idealized disk galaxy are generated using the \makedisk\ code \citep{Springel05}. We place  a dwarf-sized  disk galaxy at the center of a dark matter halo of mass $M_{\rm halo}=10^{11} \, \msun$ and virial radius $R_{\rm vir}=89 \, {\rm kpc}$, as in \citet{rosdahl15b}. The initial stellar mass of the galaxy is $2.10\times10^9 \, \msun$ with a bulge-to-total mass ratio of $f_{\rm bulge}\approx0.17$, and the total gaseous disk mass is $M_{\rm gas}=1.75\times10^9 \, \msun$. The gaseous and stellar metallicities are initially set to $Z_{\rm disc}=0.75\, {\rm Z_{\odot}}$,  where we assume $\rm Z_\odot=0.0134$ for the solar metallicity \citep{asplund09}.

The size of the simulated volume is 300 kpc on a side covered with $256^3$ root cells (level 8).  The coarse grids are adaptively refined up to a maximum level of 14, which corresponds to a spatial resolution of 18 pc.  We ensure that the Jeans length is resolved by at least 8 cells until the simulation reaches the maximum level of refinement.
Note that we first evolve the galaxy for $400\,{\rm Myr}$ to allow the galaxy to relax and to enter quasi-equilibrium, and then impose the ICM wind from one side of the box.  We define $t=0$ as the time at which the wind is first imposed. At the beginning of the ICM-ISM interaction ($t=100\,{\rm Myr}$, i.e. 100 Myr after the ICM wind is imposed), the mass of neutral, molecular, and ionized hydrogen in the galaxy, defined by the cylindrical volume of a radius of $r=10\,{\rm kpc}$ and a height of $z=\pm 3\,{\rm kpc}$ is $M_{\rm HI}=6.88\times10^8\,\msun$, $M_{\rm H_2}=2.74\times10^8\,\msun$, and $M_{\rm HII}=1.19 \times10^8\,\msun$, respectively. 

\begin{figure}
\centering 
\includegraphics[width=0.45\textwidth]{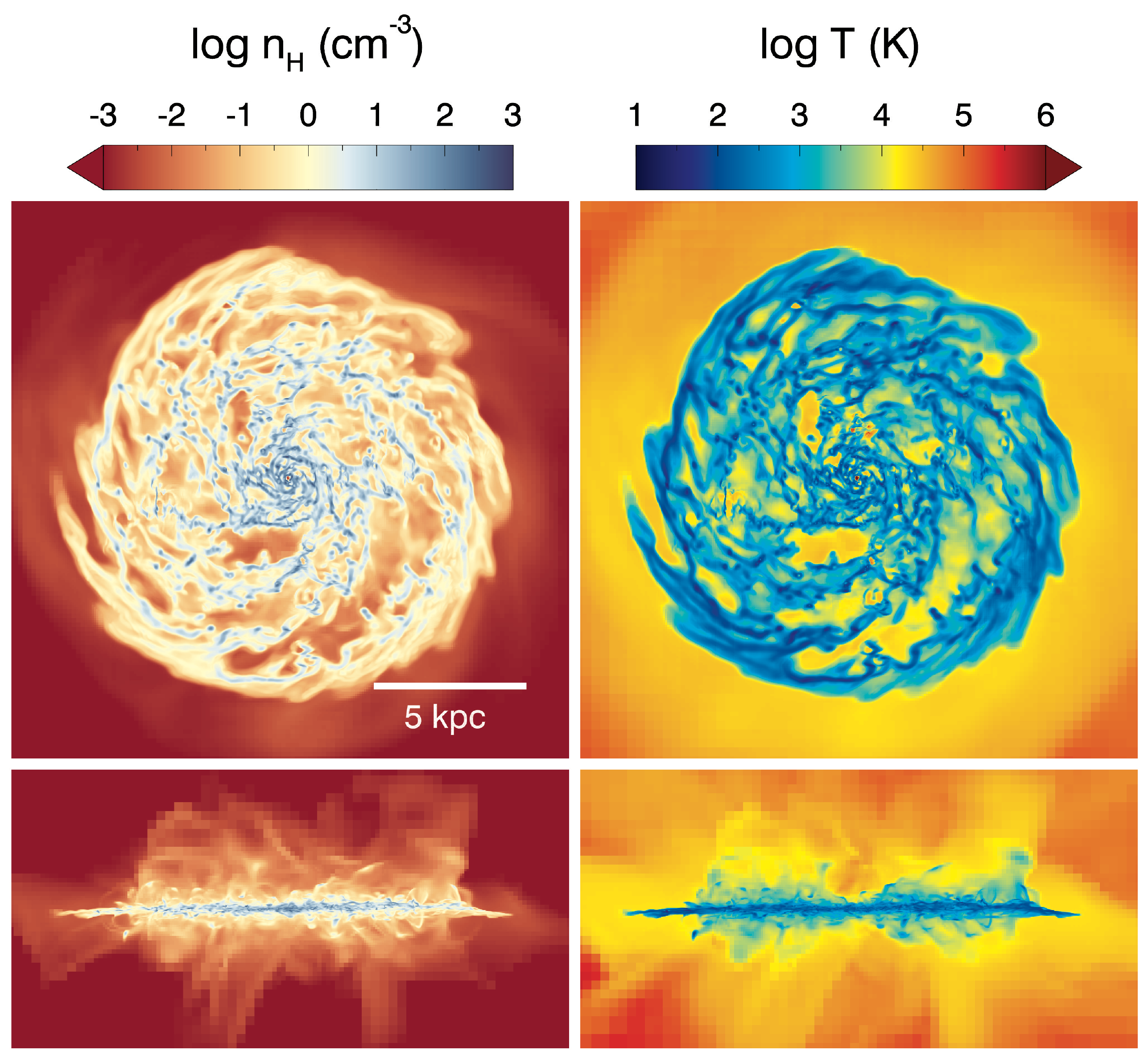}
\caption{Initial conditions of our ram pressure stripping simulations. The panels show the face-on (top) and edge-on (bottom) views of density-weighted distribution of hydrogen number density (left) and temperature (right) from the \texttt{NoWind} run at $t=100 \,{\rm Myr}$, i.e. the moment at which the simulated galaxies encounter the ICM wind in other runs. }
\label{reference}
\end{figure}

Figure~\ref{reference} shows the projected distribution of hydrogen number density ($n_{\rm H}$ in units of $\rm cm^{-3}$) and temperature (${\rm K}$) of the galaxy at $t\approx 100$ Myr. 
The scale-length $l$ of the cold gas disk (HI+H$_2$) is 1.70 kpc and its thickness $H$ is 0.12 kpc at this time. The scale-length is defined as the radius at which the density drops by $e$ compared to the one at the galactic center. The thickness $H$ is computed with $\sqrt{\int\rho z^2{\rm d}V/\int\rho {\rm d}V}$, where $\rho$ is the density of a volume ${\rm d}V$ and $z$ is the height of the volume from a galactic plane.

\subsection{ICM winds}
To mimic the interaction with the ICM in cluster environments, we impose the inflow boundary condition on one side, and outflow boundary condition on the other side. The direction of the ICM winds is set to be either face-on or edge-on, and we choose a temperature of $T_{\rm ICM}=10^7\,{\rm K}$, a metallicity of $Z_{\rm ICM}=0.004$ $\approx 0.3\,{\rm Z_{\odot}}$, and a velocity of $v_{\rm ICM}=1000~{\rm km~s^{-1}}$,  motivated by the observations of nearby clusters \citep[e.g.,][]{tormen04,hudson10,urban17}\footnote{In Appendix B, we also present the impact of slower ICM winds on star formation, which is quite similar to that of the \texttt{FaceWind} run.}. The winds first encounter the galaxy at $\sim$100 Myr after they are launched. 
Our fiducial simulations (\texttt{FaceWind} and \texttt{EdgeWind}) are run with an ICM density of $n_{\rm H,ICM}=3\times10^{-4}\cmq$, and we also examine a case with stronger ram pressure ($n_{\rm H, ICM}=3\times 10^{-3}\cmq$) for the face-on case (\texttt{FaceWind10}). The corresponding ram pressures of the winds are $P_{\rm ram}/k_{\rm B}=\rho v^{2}_{\rm ICM}/k_{\rm B}\sim5\times10^4\,{\rm cm^{-3}\,K}$ and $5\times10^5\,{\rm cm^{-3}\, K}$, respectively, which are comparable to those exerted on galaxies orbiting in the outskirts or the central regions of a cluster of mass $M_{200}\sim10^{14.8}\,\msun$ at $z=0$ \citep[][see their Figure 10]{jung18}.  We refer to the smaller ram pressure runs as moderate wind and the wind with the higher pressure as a strong wind case, hereafter.  For comparison, we run an isolated case with no wind (\texttt{NoWind}). All simulations are run for 600 Myr from the moment at which the ICM is imposed, except for the  \texttt{FaceWind10\_T6} run where we examine the effect of the temperature of the ICM wind for 300 Myr (see Table~\ref{tab:ic}).

\begin{table}
    \centering
    \caption{Simulation parameters. From left to right, the columns indicate the model name, velocity ($v_{\rm ICM}$), hydrogen number density ($n_{\rm H,ICM}$), ram pressure of the wind, and the ICM temperature. All simulations include photo-ionization heating, direct radiation pressure, photo-electric heating on dust, and mechanical supernova feedback. The maximum resolution of the simulations is $\Delta x_{\rm min}=18\,{\rm pc}$. }
    \begin{tabular}{lcccc}
    \hline
    Model  & $v_{\rm ICM}$ & $n_{\rm H,ICM}$ & $P_{\rm ram}/k_{\rm B}$ & $T_{\rm ICM}$ \\ 
       &  [\kms] & [${\rm cm^{-3}}$] & [${\rm K \,cm^{-3}}$] & [${\rm K}$] \\
       \hline
     \texttt{NoWind} & -- & -- & -- & --  \\
     \texttt{FaceWind} & $v_z=10^3$ & $ 3\times 10^{-4}$ & $5\times 10^4$  & $10^7$  \\
      \texttt{EdgeWind} & $v_x=10^3$ & $3\times 10^{-4}$ & $5\times 10^4$  & $10^7$  \\
     \texttt{FaceWind10} & $v_z=10^3$ & $3\times 10^{-3}$ & $5\times 10^5$   & $10^7$  \\
      \texttt{FaceWind10\_T6} & $v_z=10^3$ & $3\times 10^{-3}$ & $5\times 10^5$   & $10^6$  \\     
    \hline
    \end{tabular}
    \label{tab:ic}
\end{table}

\begin{figure*}
\centering 
\includegraphics[width=0.9\textwidth]{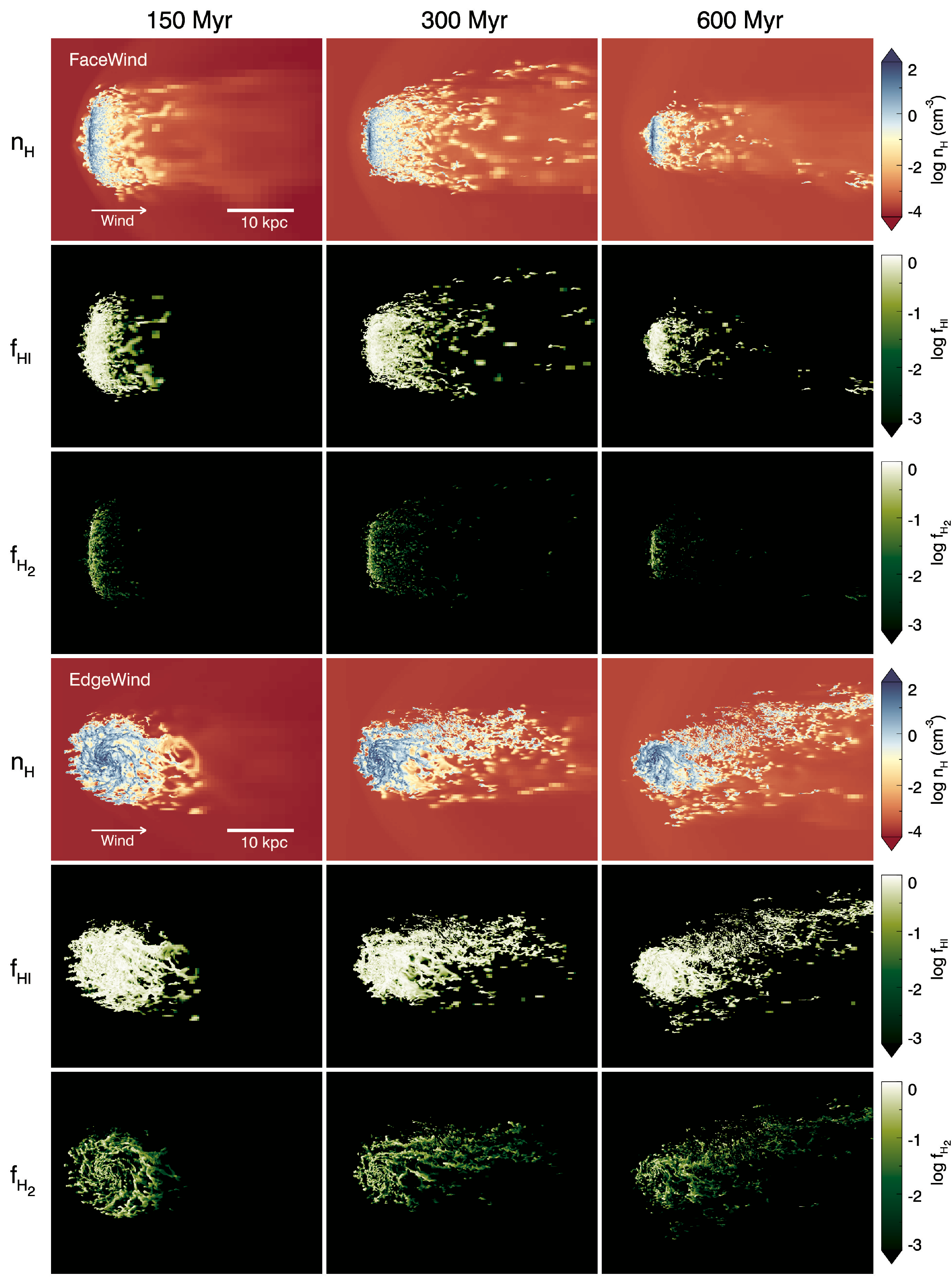}
\caption{Density-weighted projections of hydrogen number density ($\nH$), HI mass fraction ($f_{\rm HI}\equiv M_{\rm HI}/M_{\rm H}$), and H$_2$ mass fraction ($f_{\rm H2}\equiv M_{\rm H_2}/M_{\rm H}$) of the galaxy hit by the moderate ($P_{\rm ram}/k_B\sim5\times10^4{\rm cm}^{-3}\,{\rm K}$) face-on (upper three rows) and edge-on winds (bottom three rows). The white arrows indicate the direction of the wind, and the white bar measures 10 kpc. Note that the HI is much more prominent than H$_2$ in the stripped tail. The edge-on ICM wind tends to result in more extended stripped features than the face-on ICM wind. }
\label{fig:img1}
\end{figure*}

\begin{figure*}[h]
\centering 
\includegraphics[width=0.9\textwidth]{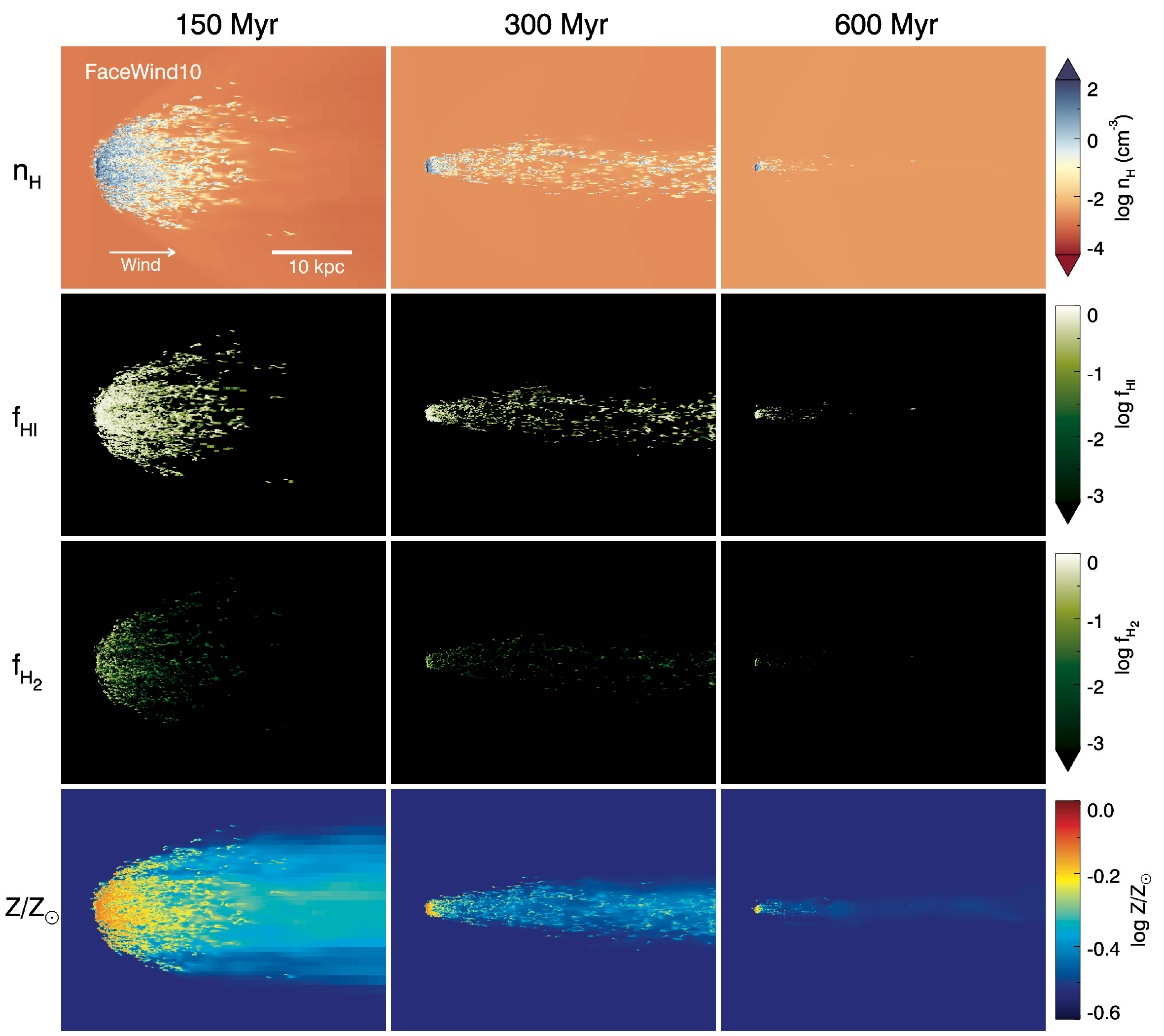}
\caption{Same as in Figure~\ref{fig:img1}, but with  an additional row illustrating projected metallicity for the run with the strong ($P_{\rm ram}/k_B\sim5\times10^5\, {\rm cm}^{-3}K$) face-on wind (\texttt{FaceWind10}). The density of the stripped gas clouds easily reaches $\nH\sim10\,\cmq$, but this is not dense enough to form stars in the extraplanar regions.  The metallicity distributions show that the stripped ISM mixes with the ICM in the tail.}
\label{fig:img2}
\end{figure*}

\section{Results}

In this section, we show how ICM winds affect the simulated gaseous disks and star formation. We also discuss the formation of HI tails and their star formation activity. 

\subsection{Ram pressure stripping in the gaseous disk}

To examine the impact of ram pressure stripping on different gas phases, we plot the distribution of hydrogen number density ($n_{\rm H}$), the fraction of neutral hydrogen ($f_{\rm HI}$), and the fraction of molecular hydrogen ($f_{\rm H_2}$)  for the \texttt{EdgeWind} and \texttt{FaceWind} runs in Figure~\ref{fig:img1}. The diffuse ISM in the galaxy, mainly composed of HI, is first swept up by the ICM winds, and stripped from the galaxy. Small clumps  with $n_{\rm H}\sim0.1$--$1\,\cmq$ (shown as yellow colors) are pushed out to $\sim 40 \,{\rm kpc}$, where they are transformed into diffuse ionized gas. The HI gas is stripped more efficiently than the H$_2$ gas, which is qualitatively consistent with observational studies  \citep[e.g][]{vollmer08,lee17} and previous numerical results that showed that dense and central gas components are not stripped easily \citep[e.g.,][]{quilis00,tonnesen09,tonnesen10}.

Figure~\ref{fig:img1} also shows that the edge-on wind generates an asymmetric tail behind the galaxy, which is different from symmetric features seen in the run with the face-on ICM wind (\texttt{FaceWind}). In the \texttt{EdgeWind} run, the gas tail is predominantly distributed toward the upper part of the image due to the interaction between the rotational motion of the disk and the wind. Gaseous components are first stripped in the region where the sum of the wind and rotational velocities are maximized, accelerated toward the $x$-direction, and then forced to flow somewhat vertically due to their initial angular momenta. During this process, some clouds that are displaced from the disk fall back to the galaxy after colliding with the stripped tail in the region $\sim$10 kpc away from the center of the galaxy. 

Unlike the runs with the moderate wind, the strong wind that cluster satellites would encounter when penetrating the central region of clusters (\texttt{FaceWind10}) removes the bulk of HI and H$_2$ gas from the galaxy rapidly (Figure \ref{fig:img2}. At  $t\ga 150\,{\rm Myr}$, even the central molecular gas component is efficiently stripped, temporarily forming a locally dense tail behind the galaxy.  The tail is composed of $\sim 8\times10^6$ cells at the level above 11 ($\Delta x < 146 \,{\rm pc}$ at 300 Myr in the \texttt{FaceWind10} run, and more than $10^6$ cells are maximally refined ($\Delta x=18\,{\rm pc}$), indicating that the strong wind induces the formation of clumpy clouds in the tail. The projected metallicity shows that the stripped ISM mixes well the ICM, enriching the metal abundance in the ICM. Further details on ram pressure stripped tails are discussed in Section 3.2.

  \begin{figure}
\centering 
\includegraphics[width=8.5cm]{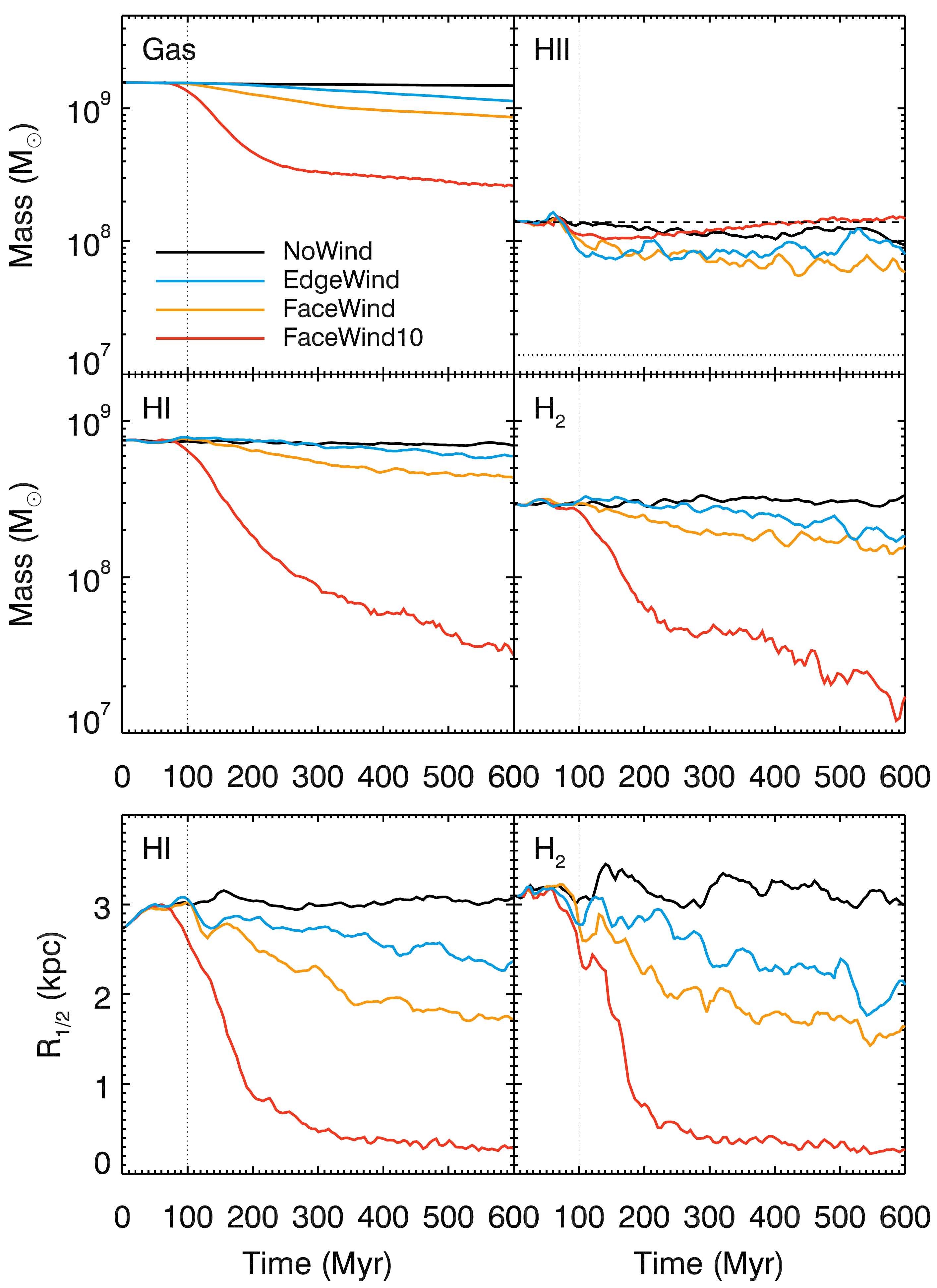}
\caption{
Top and middle rows: evolution of the gas mass in HII, HI and H$_2$ within the cylindrical volume  of radius $r=10\,{\rm kpc}$ and height $|z|= 3 \, {\rm kpc}$ centered on the galaxy. Results from runs with different ICM properties and orientation are illustrated as different color-codes, as indicated in the legend. The dashed and dotted horizontal lines in the top right panel correspond to the HII mass estimated assuming that the cylindrical volume is filled with the ICM of the strong ($n_{\rm H}=3\times10^{-3}\,\cmq$) or moderate ($n_{\rm H}=3\times10^{-4}\,\cmq$) winds, respectively. The disk gas masses in different phases decrease over time in all runs with ICM winds. Bottom row: evolution of the half-mass radii ($R_{1/2}$) of the HI and H$_2$ disks. $R_{1/2}$ gradually decreases over time in \texttt{FaceWind} and \texttt{EdgeWind}, whereas the gaseous disk in the strong wind case (\texttt{FaceWind10}) is rapidly truncated. 
}
\label{fig:mass_size}
\end{figure}

\subsubsection{Disk mass and size}

To quantify the impact of ram pressure stripping on gaseous disks, we measure the mass and half-mass radii ($R_{1/2}$) of the HI and H$_2$ disks within a cylindrical volume of radius $r=10\,{\rm kpc}$ and height $z=\pm 3\,{\rm kpc}$ centered on the galaxy in Figure~\ref{fig:mass_size}. The HI disk mass remains roughly constant in the isolated case (\texttt{NoWind}) because star formation is regulated by strong stellar feedback ($\dot{M}_{\rm star}\sim 0.2\,\msunyr$) and only $\sim10^8\,\msun$ of gas is converted into stars. The presence of the edge-on or face-on wind reduces the HI gas mass, the latter being more efficient. This can be attributed to the fact that the cross-section with the ICM wind is larger in the face-on case. As expected, the most dramatic change in HI mass is observed in the strong wind case (\texttt{FaceWind10}), where the HI and H$_2$ gas mass drops rapidly during the first $\sim 100 \,{\rm Myr}$. Furthermore the total HI and H$_2$ mass are reduced by more than an order of magnitude by $t\sim 600\,{\rm Myr}$, indicating that the strong ram pressure can indeed turn cluster satellite galaxies gas-deficient \citep{jung18}.

\begin{figure}
 \centering 
\includegraphics[width=8cm]{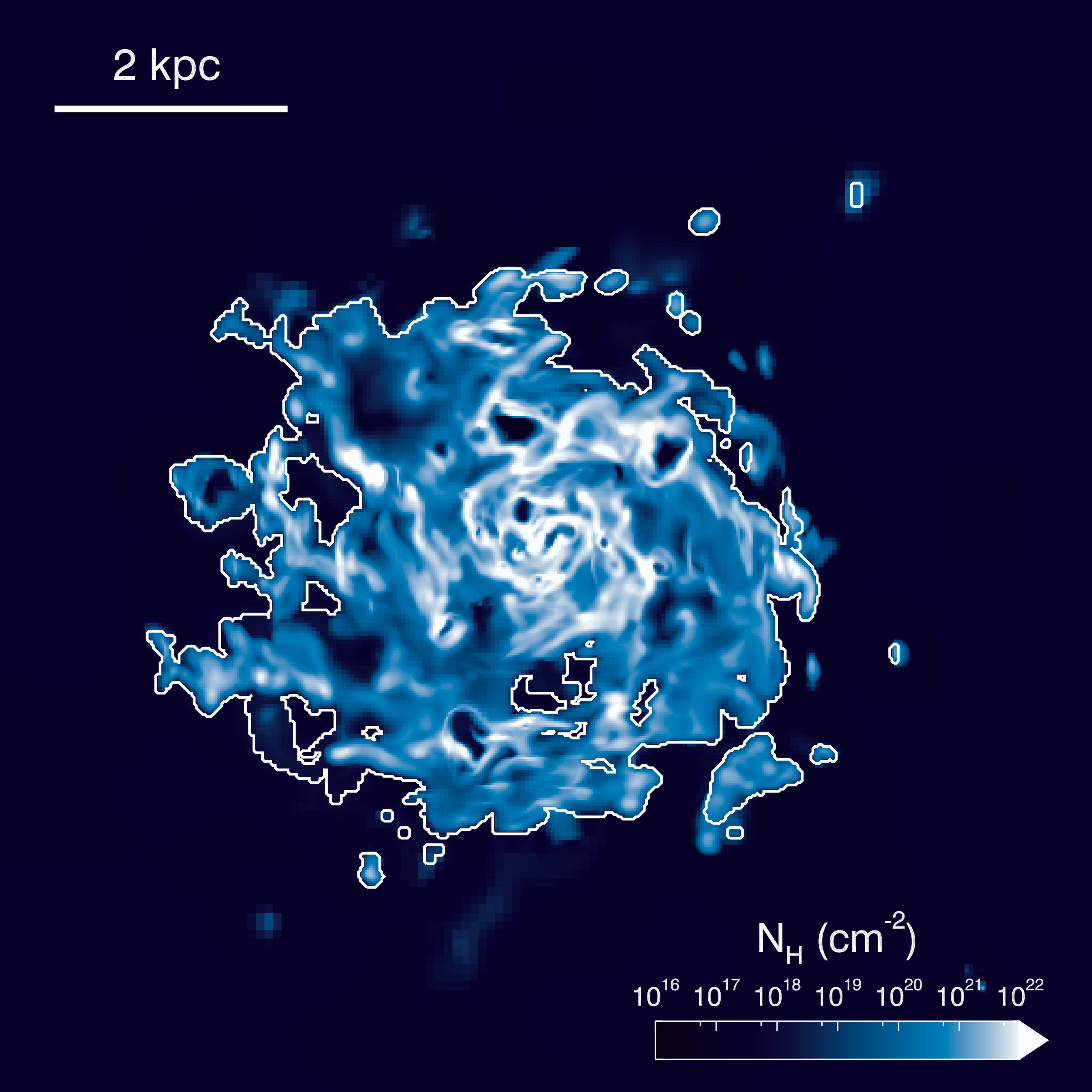}
\caption{Column density distribution of all hydrogen species (HII, HI, and H$_2$) in the central 100 pc slab ($-50\le z \le 50 \, {\rm pc}$) in the \texttt{Facewind} run at $t=600\,{\rm Myr}$. The white contours indicate the region in which the  contribution from the gas from the ICM wind to the total projected column density is less than 10 \%, which we interpret to be shielded from the ram pressure.  The fraction is estimated from the metallicity of clouds $Z$ by separating the contribution of the ISM and the ICM wind, which have the fixed metallicities of $0.75\,\rm Z_{\odot}$ and $0.3\,\rm Z_{\odot}$, respectively. The galactic disk regions are primarily shielded from ram pressure due to turbulent pressure (see  Section~\ref{sec:pturb}).}
\label{fig:gas_face}
\end{figure}

Since the outer disk is the most susceptible to the wind, the decrease in gas mass in the HI disk results in a reduction of its size (Figure~\ref{fig:mass_size}). The half-mass radius is decreased by a factor of two in the moderate face-on wind case, although the moderate edge-on wind reduces $R_{1/2}$ by $\approx 20\%$. Conversely, the majority of the extended HI disk is removed in \texttt{FaceWind10}, leaving a tiny gaseous disk with a size of $R_{1/2}\approx 0.3 \,{\rm kpc}$.
 
Not only the HI disk, but also the H$_2$ disk is significantly affected by the ram pressure. Once the molecular clumps are exposed to the ICM wind, the ram pressure subsequently removes part of the molecular gas from the disk, since it exceeds the typical pressure of the ISM (see Section 4.2). The relative differences in H$_2$ mass between \texttt{NoWind} and the runs with the ICM winds are equally notable as for HI gas mass. In particular, the reduction in H$_2$ size observed in the runs with the ICM wind,  compared with the \texttt{NoWind} case, further substantiates that the decrease in H$_2$ mass is not due to the prevention of the gas reservoir from collapsing onto the disk but a direct consequence of ram pressure stripping.

Indeed, the truncated size can be explained by the Gunn-Gott criterion \citep{gunn72} for the \texttt{FaceWind} and \texttt{FaceWind10} runs. To compute the radius $r_c$ at which the gravitational restoring force balances the ram pressure, we use the Gunn-Gott equation
\begin{equation}
\rho_{\rm ICM}v^2_{\rm ICM} = -\Sigma (r_c) \partial \Phi (r_c,z)/\partial z ,
\end{equation}
where $\Phi (r_c,z)$ is the gravitational potential measured from the gas, stars, and dark matter particles from the simulation at radius $r_c$ and a height $z$ in a cylindrical coordinate system, and $\Sigma (r_c)$ is the gas column density at radius $r_c$. The column density is obtained from the gas component located within a shell of radius $[r_c,r_c+\Delta r]$ and a height $|z|<3$ kpc, consistent with the definition of the galactic disk in this study.  We then compute the truncation radius at the disk thickness $z=H=+120\,{\rm pc}$ just after the galaxy encounters the wind ($t\sim140$ Myr), which is estimated to be  2.0 kpc and 0.8 kpc for \texttt{FaceWind} and \texttt{FaceWind10}, respectively. We confirm that these are in good agreement with the sharp break in the density profile of the simulated galaxies,   $r_c \sim 2$--$3\,{\rm kpc}$ in the \texttt{FaceWind} run or $r_c \sim 1 \,{\rm kpc}$ in the \texttt{FaceWind10} run at 600 Myr. Figure~\ref{fig:gas_face} shows the distribution of hydrogen at $t=600\,$Myr, confirming good agreement with the truncate radius in the \texttt{FaceWind} run.

\begin{figure}
\centering 
\includegraphics[width=8cm]{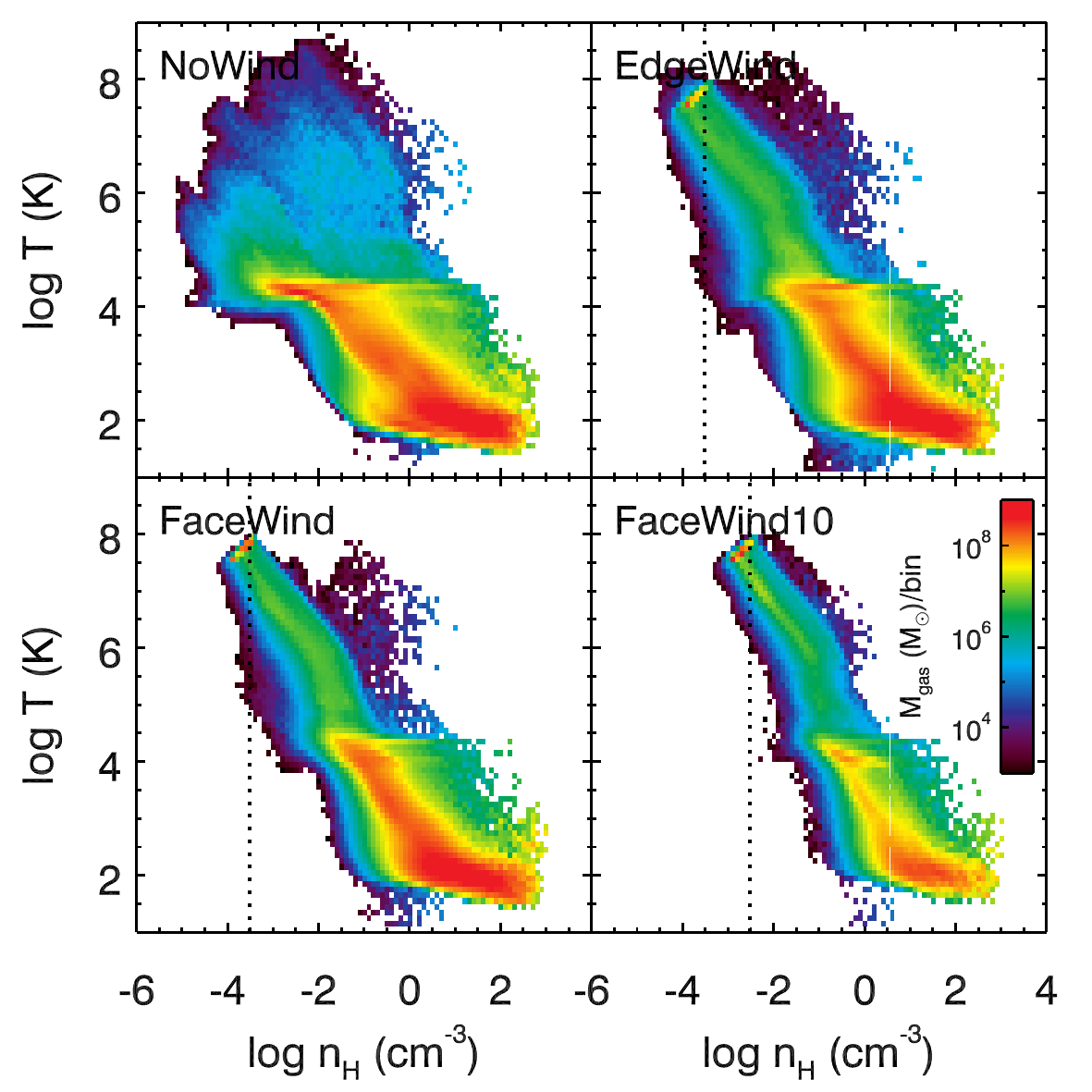}
\includegraphics[width=7.8cm]{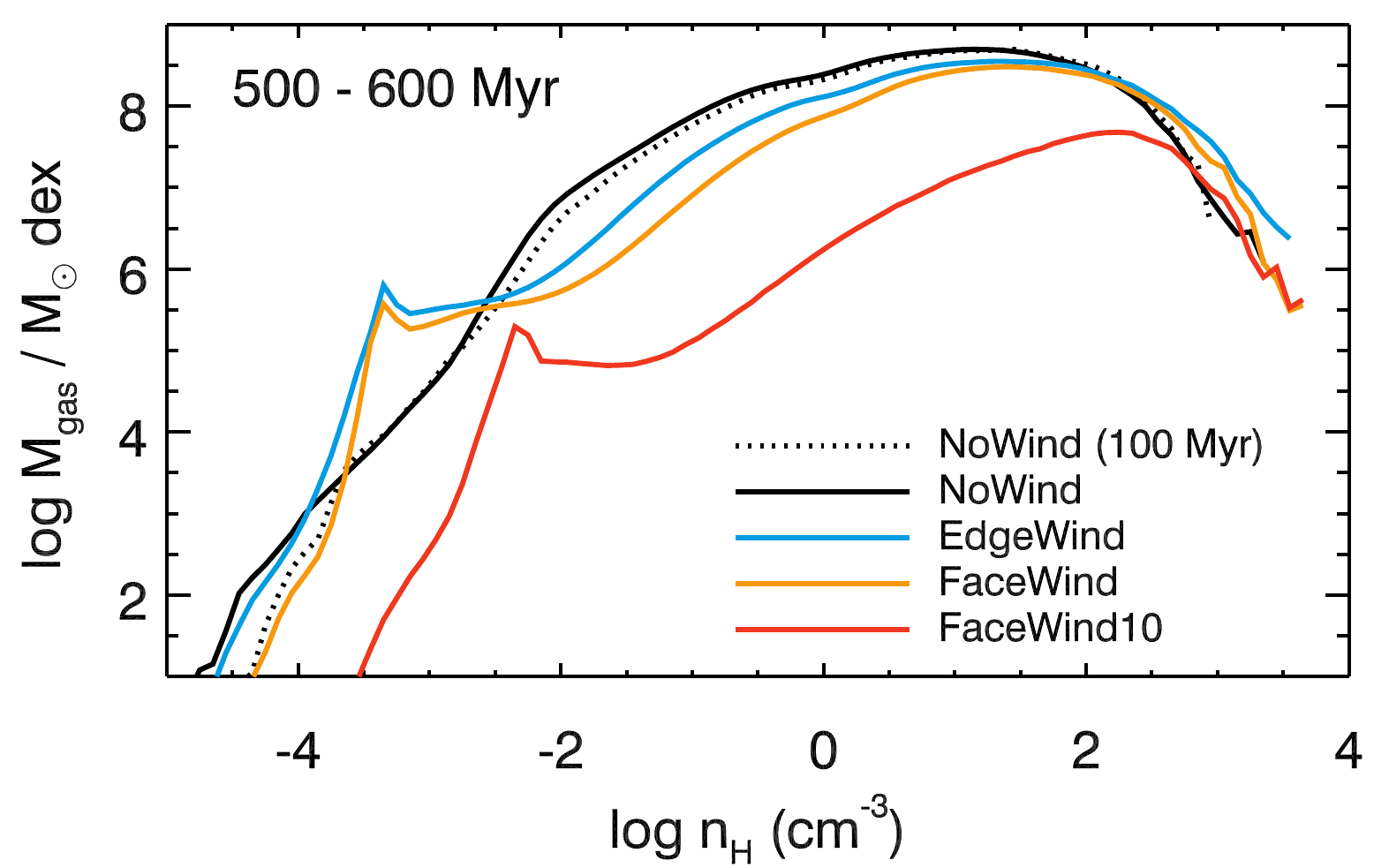}
\caption{Upper panels: gas phase diagram of the simulated galaxies affected by different ICM winds. The top left panel displays the distribution in the \texttt{NoWind} run at 100 Myr, i.e., just before the wind hits the galaxy, and the other three panels correspond to the phase diagram at 600 Myr. We note that the \texttt{NoWind} run is not shown here at 600 Myr because only negligible evolution is observed in the phase space. Different colors indicate the gas mass shown in each bin, as indicated in the legend. The density of the ICM winds are shown as vertical dotted lines.  The green plume around $T\sim10^6\,{\rm K}$ forms from the ICM gas that interacts with the ISM and cools. Lower panel: the averaged distribution of gas mass measured within the cylindrical volume around the galaxy at $500\le t \le 600 \, {\rm Myr}$. The ICM winds preferentially remove the diffuse gas component, making the gas distribution in the phase diagram narrower.}
\label{fig:gas_phase}
\end{figure}

\subsubsection{Temperature and density distribution}

Different ICM winds shape galaxy evolution distinctly, resulting in a different density and temperature distributions of a gaseous disk. In Figure~\ref{fig:gas_phase}, we present the density and temperature distribution in the cylindrical volume enclosing the galaxy (i.e. $r\le 10 \,{\rm kpc}$ and $|z| \le 3 \,{\rm kpc}$)  at $t=100\,{\rm Myr}$ for the \texttt{NoWind} run and $t=600\,{\rm Myr}$ for the runs with the ICM winds, for comparison. The ICM is kept hot and diffuse until it encounters the galaxy. Once the ICM wind interacts and mixes with the ISM, it cools and collapses behind the galaxy, leaving a notable feature at $10^5\la T \la10^7\,{\rm K}$ (green plume).   

We find that ram pressure blows away a significant fraction of the diffuse ISM with densities lower than $\rho_{\rm ICM}$. In contrast, the dense ISM with $\nH\ga 1\,\cmq$ is self-shielded from the wind and remains mostly intact, leading to a narrow gas distribution towards high densities in the phase diagram. By contrast, the presence of a strong ICM wind destroys massive clouds with $\nH\sim100\,\cmq$. 

The difference in the distributions of gas mass is illustrated more clearly in the bottom panels of Figure~\ref{fig:gas_phase}, where we plot the gas density distribution at 600 Myr within the cylindrical volume of radius $r<3 \,l_{\rm s}$ and height $|z|< 2H$, where $l_{\rm s}$ is the scale-length and $H$ is the thickness of the disk. When isolated, the simulated galaxy shows minimal evolution in the mass distribution with time (black dotted vs solid line). However, a prominent peak develops around $\rho\sim\rho_{\rm ICM}$ once the ICM wind reaches the disk. The gas mass in the two runs with the moderate wind (\texttt{EdgeWind} and \texttt{FaceWind}) are enhanced at high densities ($ n_{\rm H}\ga100\cmq$), compared to those in the isolated case (\texttt{NoWind}), which suggests that star formation can be enhanced in \texttt{EdgeWind} and \texttt{FaceWind} than in \texttt{NoWind} (as we confirm later in Section~\ref{sec:star_formation}). In addition, the majority of the gas disk is stripped in the presence of the strong ICM wind.

\begin{figure*}
\centering 
\includegraphics[width=16cm]{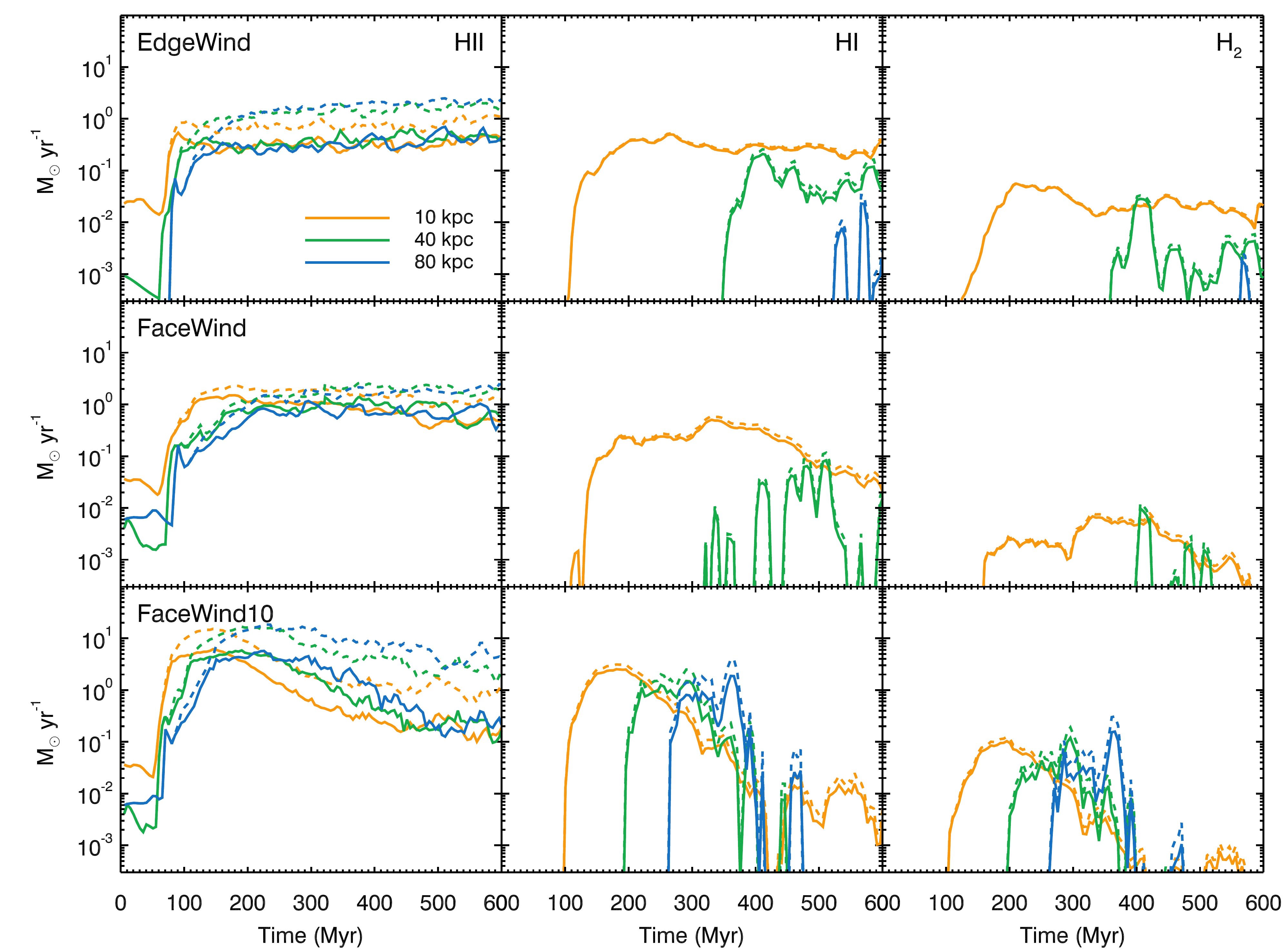}
\caption{Outflow fluxes measured along the direction of the ICM wind within a circular surface of radius $r=10\,{\rm kpc}$ placed at a distance 10, 40, and 80 kpc away from the galaxy center. From top to bottom, each row indicates the flux measured in the \texttt{EdgeWind}, \texttt{FaceWind}, and \texttt{FaceWind10} runs, respectively. The fluxes of different phases (HII, HI and H$_2$) are shown in the left, middle, and right columns, respectively. The outflow fluxes originating from the simulated galaxy with $Z=0.01=0.75\,{\rm Z_{\odot}}$ are shown as solid lines, whereas fluxes including the gas cooled or added from the ICM, are shown as dashed lines. Most neutral gas (HI and H$_2$) is blown directly from the galactic disk, and only $\sim20\%$ of the neutral gas in the tail is cooled from the ICM in the two runs with the moderate winds. In contrast, the strong wind induces efficient mixing, and up to $\sim60\%$ of the neutral hydrogen in the tail is cooled from the ICM. 
}
\label{fig:outflow_rate}
\end{figure*}

\begin{figure*}
\centering 
\includegraphics[width=16cm]{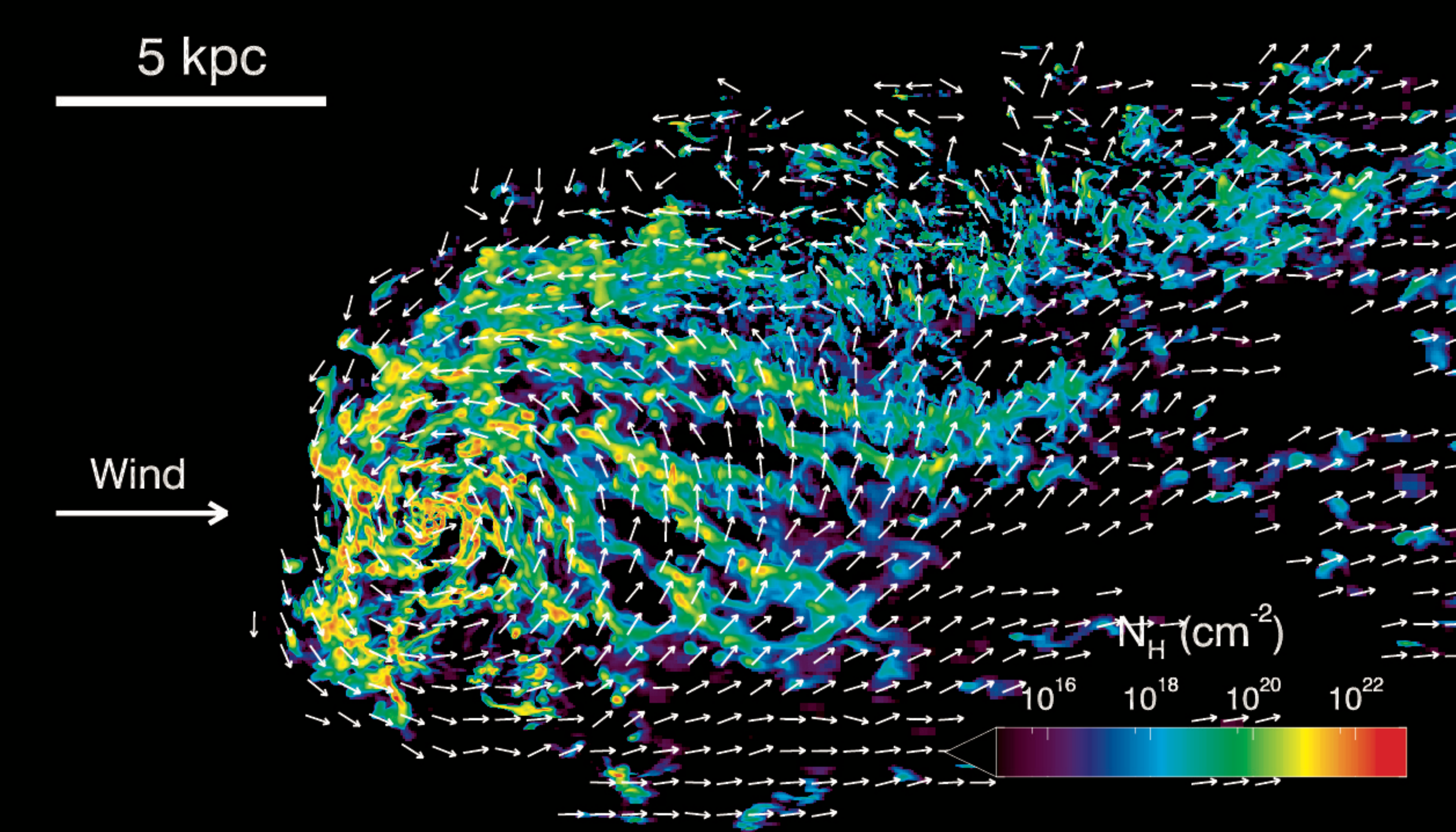}
\caption{Velocity field of the simulated galaxy in the \texttt{EdgeWind} run at $t=376 \,{\rm  Myr}$. The color denotes the column density of molecular hydrogen. Notice that some of the gas clouds at $r_c\sim 10\,{\rm kpc}$ fall back after being blown out by the wind.}
\label{fig:EdgeWind_velocity}
\end{figure*}

\subsection{Formation of the ram pressure stripped tail}

Galaxies moving inside the ICM form characteristic ram pressure stripped tails. In this section, we investigate the formation of the tail by measuring the outflowing flux from the disk and the evolution of tail mass in the runs with different ICM winds. 

We define the tail as gas components located inside a cylindrical volume of a radius $r \le 10\,{\rm kpc}$ and a height $z>10\, {\rm kpc}$ (or $x>10\,{\rm kpc}$ for the \texttt{EdgeWind} run) along the wind direction. Note that we only consider gas with $Z>0.004$ so that the pure ICM wind is not included. We also use the different metallicities of the ISM ($Z_{\rm ISM}=0.01$) and ICM ($Z_{\rm ICM}=0.004$) to distinguish their relative contributions to the formation of the tail. 
The outflow rates are measured from gas crossing the circular surface of radius of $r=10\,{\rm kpc}$ at 10, 40, and 80 kpc from the galaxy center.  Here we define the galaxy center as the center of stellar mass. We do not separate the contribution from galactic outflows due to stellar feedback, as it is negligible in these regions ($\dot{M}_{\rm out}\ll 0.01 \,\msunyr$), compared to the flux due to ram pressure.

Figure~\ref{fig:outflow_rate} shows that the majority of the gas that leaves the \texttt{FaceWind} galaxy is in the form of ionized hydrogen (second rows). The HII stripping rates ($0.3-1\,\msunyr$, solid lines) are a factor of few higher than the star formation rates of the \texttt{NoWind} run ($\dot{M}_{\star}\sim0.1\,\msunyr$). 
The cumulative HII, HI, and H$_2$ masses crossing the surface at $10\,{\rm kpc}$ during $0<t<600 \,{\rm Myr}$ are estimated to be $4.4\times10^{8}$,  $0.9\times10^{8}$, and $10^6\,\msun$, respectively. However, we find that the decrease in HII mass from the galaxy ($0.5\times10^{8}\,\msun$) is much lower than the cumulative HII mass crossing the surface, which indicates that a substantial fraction of the HII flux measured at 10 kpc originated from the neutral phase. Indeed, the decrease in HI mass ($\sim4\times10^{8}\,\msun$) is higher than the total HI mass crossing the surface ($0.9\times10^{8}\,\msun$), as seen in Figure~\ref{fig:mass_size}, supporting the view that HI is stripped and then collisionally ionized due to the ICM wind before it leaves the galaxy. 

The HI flux increases rapidly up to $\sim 0.5\,\msunyr$ at $10\, {\rm kpc}$ in \texttt{FaceWind}, but declines over time as a diffuse ISM is removed from the galaxy.  Although $\sim 30\%$  of the HI flux measured at 40 kpc turns out to have cooled from the ICM wind, the HI gas is continuously ionized, as it moves away from the galaxy. Eventually, little HI gas persists in the case with the moderate face-on wind.

In the case of \texttt{EdgeWind}, the stripping process seems somewhat different.  The outflow flux measured at 10 kpc at $t\approx 100\, {\rm Myr}$ increases more rapidly than in the face-on case, since part of the galactic disk is already rotating along the same direction of the ICM, as illustrated in Figure~\ref{fig:EdgeWind_velocity}. Furthermore, the diffuse medium above the galactic mid-plane is directly exposed to the ICM wind and unimpeded by the ISM, contrary to the \texttt{FaceWind} case. However, the total amount of the gas that is directly stripped from the galaxy and crosses the 10 kpc surface (orange solid lines) is actually lower than in the face-on wind case by 60\% (see also Figure~\ref{fig:mass_size}), and the stripping rate is only comparable to that of star formation in the  \texttt{NoWind} run. 

Despite the less significant ram pressure stripping, we find that $\sim 30\%$ more gas leaves the galaxy and is measured at 10 kpc  in the form of HI from the \texttt{EdgeWind} run, compared with the \texttt{FaceWind} case. This is also potentially due to the coherent acceleration of the HI gas along the direction of the rotation in the galactic disk. Similarly, a larger amount of stripped molecular hydrogen is observed at $40\, {\rm kpc}$ in the \texttt{EdgeWind} run than in the \texttt{FaceWind} run.  On average, the total H$_2$ mass present in the tail region at $300<t<600\,{\rm Myr}$ is $\sim 10^7\,\msun$ in the \texttt{EdgeWind} run, whereas the mass is an order of magnitude smaller in the \texttt{FaceWind} run. Similarly, as in the \texttt{FaceWind} run, once blown out from the galaxy, the HI gas is continuously ionized, and a tiny amount ($4\times10^5\,\msun$) of gas travels out to 80 kpc.

We find that molecular hydrogen is inefficiently carried away by the moderate winds. The cumulative amount of H$_2$ passing through the 10 kpc surface is  $\sim10^6$ or $10^7\,\msun$ for the \texttt{FaceWind} and \texttt{EdgeWind} runs, which is less than 10\% of the cumulative HI flux. The H$_2$ stream at 40 kpc is visible, but its flux is even lower, indicating that H$_2$ clouds are destroyed and/or dissociated by interactions with the moderate ICM wind. Although the ICM wind material condenses to form additional molecular hydrogen at the skin of the stripped molecular clouds, this does not account for a large fraction of the molecular hydrogen in the tail ($<20\%$) in the two runs with the moderate winds. 

Although not included in Figure~\ref{fig:outflow_rate}, we also confirm that the galactic rotation can enhance inflowing flux when the simulated galaxy encounters the edge-on ICM wind. The velocity field presented in Figure~\ref{fig:EdgeWind_velocity} shows that asymmetric velocity structures develop at $r\ga 5 \,{\rm kpc}$ where the gas clouds accelerated by galactic rotation collide with the ram pressure stripped tail. In addition, a significant fraction of the gas tail in the upper region falls back to the galaxy, with an infall rate of $0.01$-$0.02\,\msunyr$, which is more significant than that in the \texttt{NoWind} or \texttt{FaceWind} run ($<0.001\, \msunyr$).

Finally, a large amount of the ISM is blown away from the galaxy hit by the strong face-on wind, showing a maximum stripping rate of $\sim5\, \msunyr$ at all three distances (Figure~\ref{fig:outflow_rate}). However, the fluxes decrease to a level similar to that in the moderate wind runs, once the majority of the galactic gas is stripped within a short timescale of $\sim 100\,{\rm Myr}$ and little diffuse ISM remains in the galaxy \citep[Figure~\ref{fig:mass_size}, see also][]{tonnesen19}. 
Because of the large amount of stripped material, the mixing of the ISM and ICM occurs more efficiently under the strong wind as well. Out of the total HI mass in the tail at $300 \, {\rm Myr}$,  43.2\% of HI is cooled from mixing with the ICM. For comparison, only 14.9\% and 7.4\% of HI are cooled from the ICM in the \texttt{FaceWind} and \texttt{EdgeWind} runs, respectively. We note, however, that the exact amount of cooled gas may depend on the assumed temperature of the ICM, as we will discuss in Section~\ref{sec:strong_tail}.

\begin{figure}
\centering 
\includegraphics[width=8.5cm]{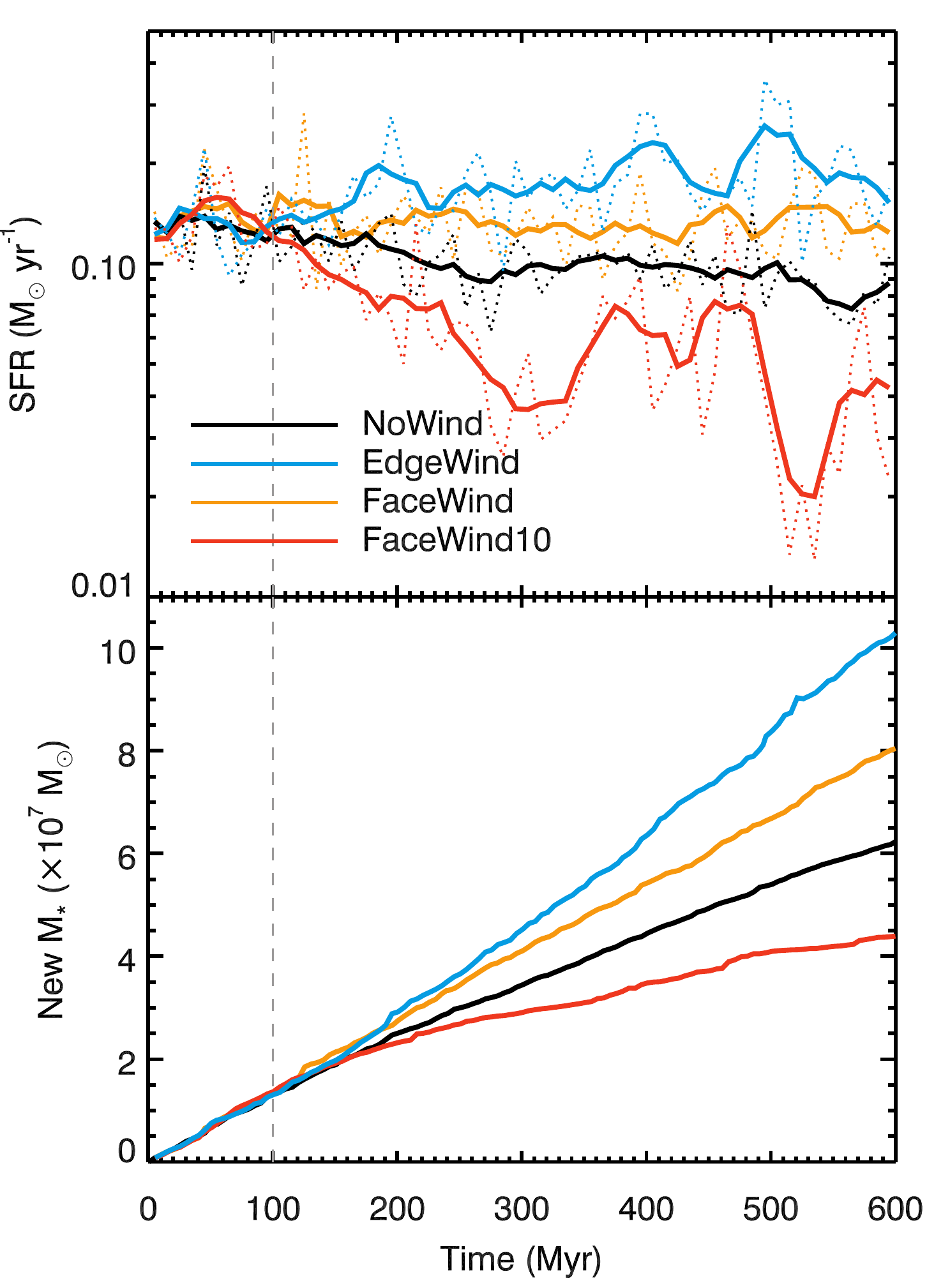}
\caption{Evolution of the star formation rates (top) and the cumulative stellar mass formed (bottom). The vertical dashed lines mark the epoch when the ICM wind hits the gaseous disk ($t\sim100\,$Myr).  In the top panel, the solid and dotted lines present SFRs averaged over 50 Myr and 10 Myr, respectively. The presence of the moderate wind enhances the star formation rates by $\sim40$--$80\%$, compared to the case in the \texttt{NoWind} run. Conversely, the star formation rates are reduced by a factor of $\sim 2$ when the strong ICM wind is imposed.}
\label{fig:sfr}
\end{figure}

\subsection{Star formation}
\label{sec:star_formation}

Ram pressure stripping is believed to eventually quench star formation by removing cold gas reservoirs from cluster satellite galaxies. However, the results of observational studies suggest that ram pressure may trigger or even enhance star formation on short timescales \citep{crowl06,crowl08,merluzzi13,scott13,kenney14,lee17,vulcani18}.
In this subsection, we investigate the dual effects of ram pressure on star formation activities based on our simulations.

\subsubsection{Star formation in the galactic disk}

We find that star formation is enhanced in the runs with the moderate ICM wind, when compared to the \texttt{NoWind} case (Figure~\ref{fig:sfr}). Although the total gas mass decreases by a factor of two (Figure~\ref{fig:mass_size}), the total stellar mass formed between $100<t<600 \,{\rm Myr}$ in the \texttt{FaceWind} run increases by $\sim 40\%$, compared to the \texttt{NoWind} run. The increase in star formation turns out to be even more notable in the \texttt{EdgeWind} ($\sim80\%$) case, as the amount of high-density gas increases (see Figure~\ref{fig:gas_phase}). We also examine whether or not the star formation efficiency per free-fall time is enhanced, compared to the \texttt{NoWind} case.  We find that when the criteria for star formation are met, the average star formation efficiency is similar for the two runs ($\epsilon_{\rm ff}\approx0.21$) (see Equation 2), as the moderate wind cannot penetrate the ISM and influence the dynamics of the star-forming clouds directly. Conversely, the strong face-on wind suppresses star formation by a factor of a few, especially in the late stage of the ICM-ISM interaction. Not only the total amount of star-forming gas with $\nH\ga100\,\cmq$ decreases, but $\epsilon_{\rm ff}$ also decreases slightly to 0.19, indicating that the clouds are less gravitationally bound due to the turbulent interaction with the ICM wind. However, the reduction in the star formation rates does not appear as significant as the change in the molecular or HI mass in  \texttt{FaceWind10} at $t=600 \,{\rm Myr}$ (Figure~\ref{fig:mass_size}), as dense star-forming regions are the least affected by the wind (see Figure~\ref{fig:gas_phase}).

\begin{figure}
\centering 
\includegraphics[width=8.0cm]{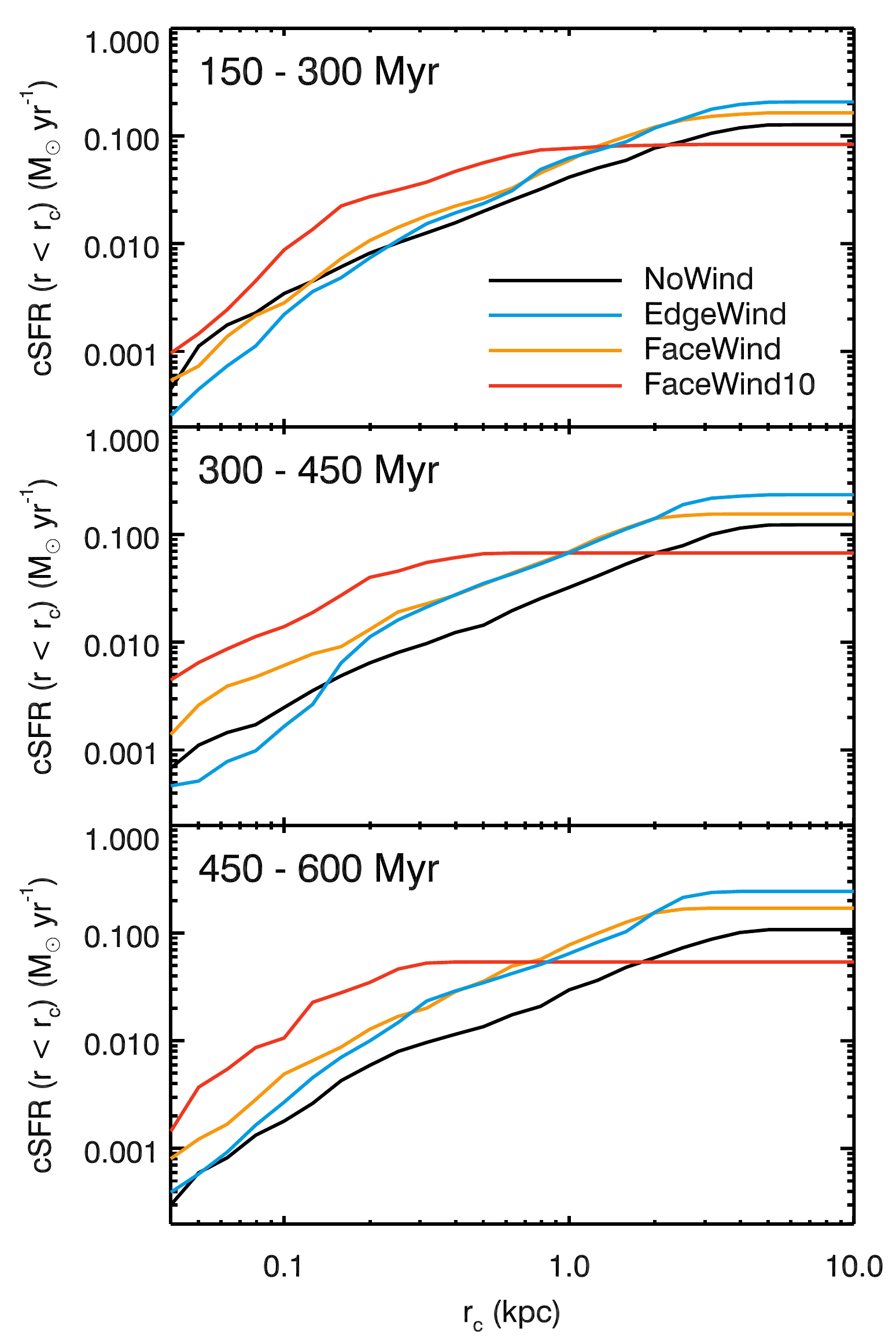}
\caption{Cumulative star formation rates (i.e. within the given radius) as a function of the cylindrical radius ($r_c$), averaged over 150 Myr. Results from different models are shown as different colors, as indicated in the legend. Star formation activities in \texttt{FaceWind} are enhanced in the central region ($r_c\la 2\,{\rm kpc}$), but suppressed at $r_c\ga 2\,{\rm kpc}$, compared to the \texttt{NoWind} run (black lines). Similarly, the strong face-on wind shows the dual effects on star formation, but with a smaller truncation radius at $r_c\sim 0.3\,{\rm kpc}$. In contrast, star formation is boosted preferentially at  $r_c\sim0.5-2\,{\rm kpc}$ in the \texttt{EdgeWind} run.}
\label{fig:csfr_evo}
\end{figure}

\begin{figure*}
\centering 
\includegraphics[width=0.9\textwidth]{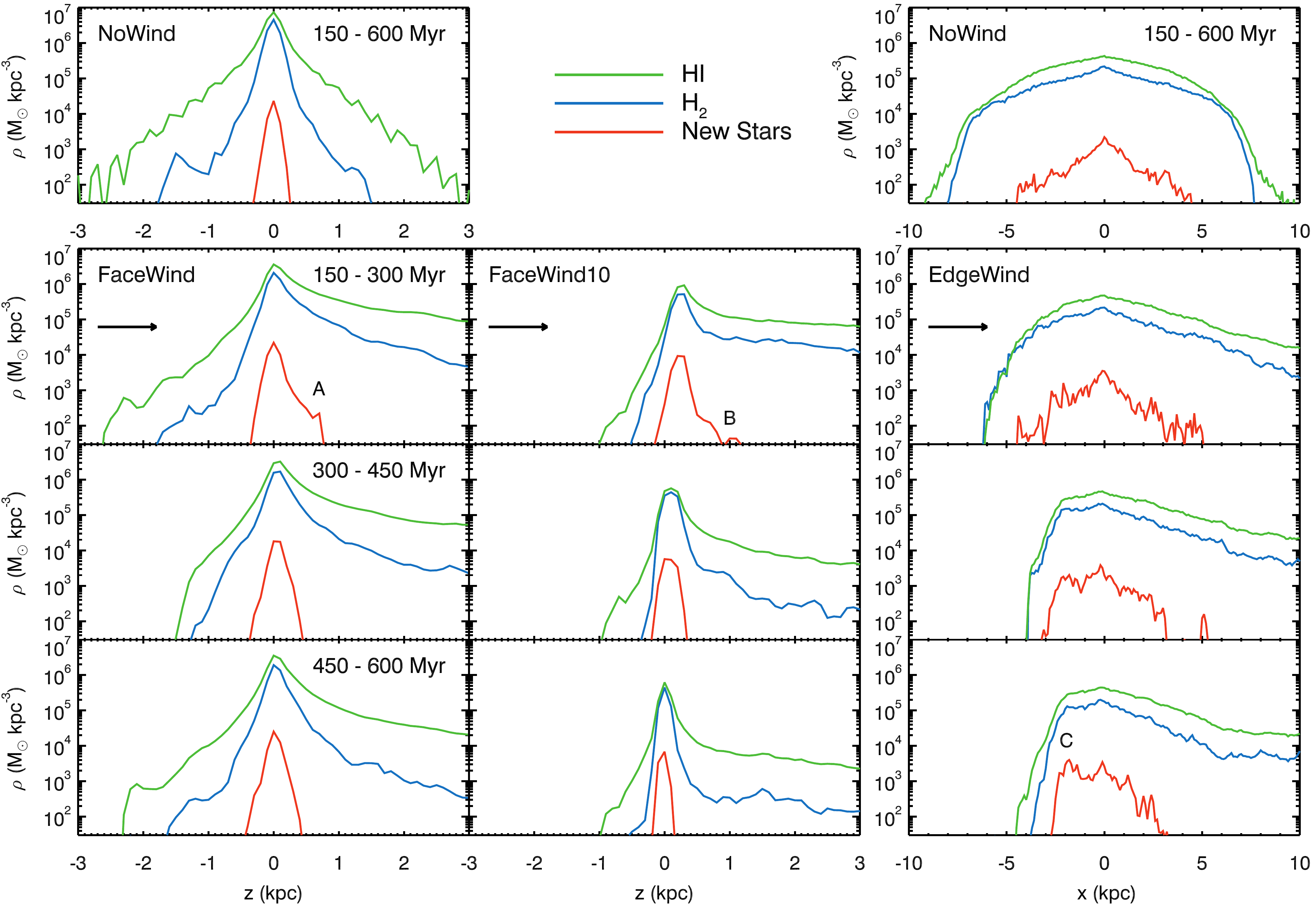}
\caption{Densities of HI (green), H$_2$ (blue), and star particles younger than $10 \, {\rm Myr}$ (red), measured along the direction of the wind. We use the center of stellar mass to define $z=0$ and $x=0$. The run without the ICM wind is shown in the top rows, while the runs with face-on or edge-on wind are displayed in the lower panels. The arrow in each panel indicates the direction of the wind.  Off-plane star formation sites are marked as \texttt{A} and \texttt{B}. The edge-on wind compresses the gaseous disks, significantly boosting star formation at the interface between the ICM wind and the ISM (marked as \texttt{C} in the bottom right panel).}
\label{fig:sfr_face}
\end{figure*}

The enhanced star formation pattern in the runs with the moderate winds is analyzed further in Figure~\ref{fig:csfr_evo}. We compute the the cumulative star formation rate (cSFR) profile in each snapshot ($\Delta t\sim5$ Myr) as a function of radius, and plot the average over 150 Myr ($\sim$30 snapshots).  We find that newly formed stars are distributed differently depending on the direction of the wind. The star formation rates in the runs with the face-on wind are enhanced in the central region ($r_c \la 2 {\rm kpc}$) of the galaxy, compared to the \texttt{NoWind} case. The trend is more pronounced in the later stage of the evolution ($t>300\,{\rm Myr}$) because the ISM that initially protects the star-forming regions from the ICM wind is stripped over time. Once a significant amount of gas is lost in the outer region $r_c \ga 2\, {\rm kpc}$ of the \texttt{FaceWind} disk, star formation becomes  suppressed.  However, the enhancement in star formation seems to occur only in the very central region ($r_c \la 0.5 {\rm kpc}$) in the \texttt{FaceWind10} run because it is directly exposed to the ICM wind due to efficient stripping of the outer gaseous disk.  On the contrary, the edge-on wind triggers star formation preferentially in the outer region of the galaxy,  especially in the interface between the ISM and the ICM. 

In order to understand the change in star formation, we plot the density of newly formed stars, HI, and H$_2$ along the direction of the ICM wind in Figure~\ref{fig:sfr_face}. We clearly observe that the star formation, as well as the average molecular hydrogen densities, are elevated at the interface between the ISM and the edge-on wind  (see the reddish regions in Figure~\ref{fig:EdgeWind_velocity}).  For the \texttt{EdgeWind} galaxy, the interface forms within the galaxy at $-5 \la  x \la -3\,{\rm kpc}$, which is seen as a boost in star formation at $x\sim-3\,$kpc (marked as ``C'' in the bottom right panel of Figure~\ref{fig:sfr_face}). Conversely, star formation in the opposite side ($x\ga 3\,{\rm kpc}$) of the \texttt{EdgeWind} galaxy is quenched,  mainly because the disk gas is shielded and not experiencing ram pressure.

\subsubsection{Star formation outside the galaxy}

Recent observations reveal young stars as well as dense CO clouds in ram-pressure stripped tails at several kpc from the cluster satellite galaxy \citep[e.g.,][]{poggianti17,lee17,jachym19}. We find that such extraplanar cold clouds form in the run with the strong wind during the early stage ($\sim$150 Myr) of the ICM-ISM interaction. In the top panel of Figure~\ref{fig:img2}, the gas with blue colors, corresponding to the total surface density of $N_{\rm HI}\sim 3\times 10^{21} \, {\rm cm^{-2}}$ or $30 \,{\rm \msun\, pc^{-2}}$, can easily be found at $\la 10\,{\rm kpc}$ from the galaxy. 
Similarly, dense clouds are also found in \texttt{EdgeWind} at $5$--$10 \,{\rm kpc}$ where the ICM wind interacts with the ISM in the opposite side of the interface (see the upper right area of Figure~\ref{fig:EdgeWind_velocity}). The metallicity of the dense clouds (yellow colors) is $Z\approx0.75\,{\rm Z_{\odot}}$, indicating that they originate from the ISM, and are not  mixed/cooled from the ICM. An exception to the efficient stripping is the \texttt{FaceWind} case where only the diffuse ISM is stripped mildly and does not trigger significant gas collapse.

However, no clear signature of extraplanar star formation is found in our simulations. The \texttt{FaceWind10} run does show extraplanar star formation briefly during the initial phase of the ICM-ISM interaction (SFR $\sim9.5\times10^{-3}\,\msun\,{\rm yr^{-1}}$, marked as ``B'' in Figure~\ref{fig:sfr_face}) when the ram pressure directly pushes pre-existing star-forming clouds. The \texttt{FaceWind} run also exhibits a feature similar to the SFR of $1.5\times10^{-2}\,\msunyr$ (marked as ``A'' in Figure~\ref{fig:sfr_face}). Yet, the enhanced star formation in the extraplanar regions does not last long, and we find that the young stars fall back quickly to the galactic disk, resulting in slightly extended star formation up to 1 kpc from the mid-plane of the disk. Strong stellar feedback periodically induces off-plane ISM winds, puffing HI and H$_2$ out from the disks, and this is particularly visible in the last stage (450-600 Myr) in \texttt{FaceWind} run. In the case of \texttt{EdgeWind}, some the clouds are displaced up to $x\sim 5\,{\rm kpc}$ at $150<t<450\, {\rm Myr}$, forming a small amount of stars, but in general their volumetric densities turns out to be not high ($\nH\sim1-10\,{\rm cm^{-3}}$) enough to trigger star formation in the extraplanar cold clouds.

\section{Discussion}

In this section, we discuss the complicated effects of ram pressure on star formation in the disk and tails, the potential role of turbulence against ram pressure, and limitations of this study.  

\subsection{Does ram pressure quench or trigger star formation?}

Previous studies have revealed the evident impact of ram-pressure stripping on HI disks in many cluster satellites \citep{davies73,giovanelli85,giraud86,warmels88b,cayatte90,scodeggio93,chung09}, but there has been a long debate on the stripping of the molecular disk \citep{stark86,kenney89,rengarajan92, crowl05,vollmer08,abramson14,boselli14,kenney15,lee17}. A notable feature of ram-pressure-stripped galaxies is that star formation seems to be enhanced before quenching, compared with similar galaxies in less dense environments \citep{poggianti04,crowl06,ma08,poggianti09,merluzzi13,kenney14,lee17,vulcani18}. This suggests that ram pressure stripping may play a more complex role than simply removing cold gas and suppressing star formation.

Our experiments confirm that if a newly accreted galaxy experiences mild ram pressure in the cluster outskirts (\texttt{FaceWind}, \texttt{EdgeWind}), the star formation can be temporarily boosted on a timescale of several hundred Myr. We confirm that this is still the case even when we apply a ram pressure that is 10 times smaller than that of the run with the moderate face-on wind (\texttt{FaceWind\_slow}, see Appendix B). This is qualitatively consistent with previous numerical studies  \citep{schulz01,vollmer01,bekki03,kronberger08,kapferer08, kapferer09, steinhauser12}. We find that this continues during the entire course of our simulations ($\sim 500 \,{\rm Myr}$), which is a sizable fraction of the typical orbital timescale of cluster satellite galaxies \citep[e.g.,][]{khochfar06}. Even when they enter the central region of the cluster (\texttt{FaceWind10}), the star formation is likely to be locally and temporarily boosted, as shown in Figure~\ref{fig:csfr_evo}. Star formation is immediately suppressed if strong ram pressure is exerted (see Figure~\ref{fig:sfr}) or applied in the outer disk where the low-column density gas is significantly removed (Figure~\ref{fig:csfr_evo}). This aptly demonstrates that both quenching and triggering can operate on galaxies falling into massive clusters, reinforcing the view that the reduction in star formation is closely linked to the peri-center distance and the number of peri-center passages of the cluster satellites \citep[e.g.,][]{rhee20}.

\begin{figure*}
\centering 
\includegraphics[width=16cm]{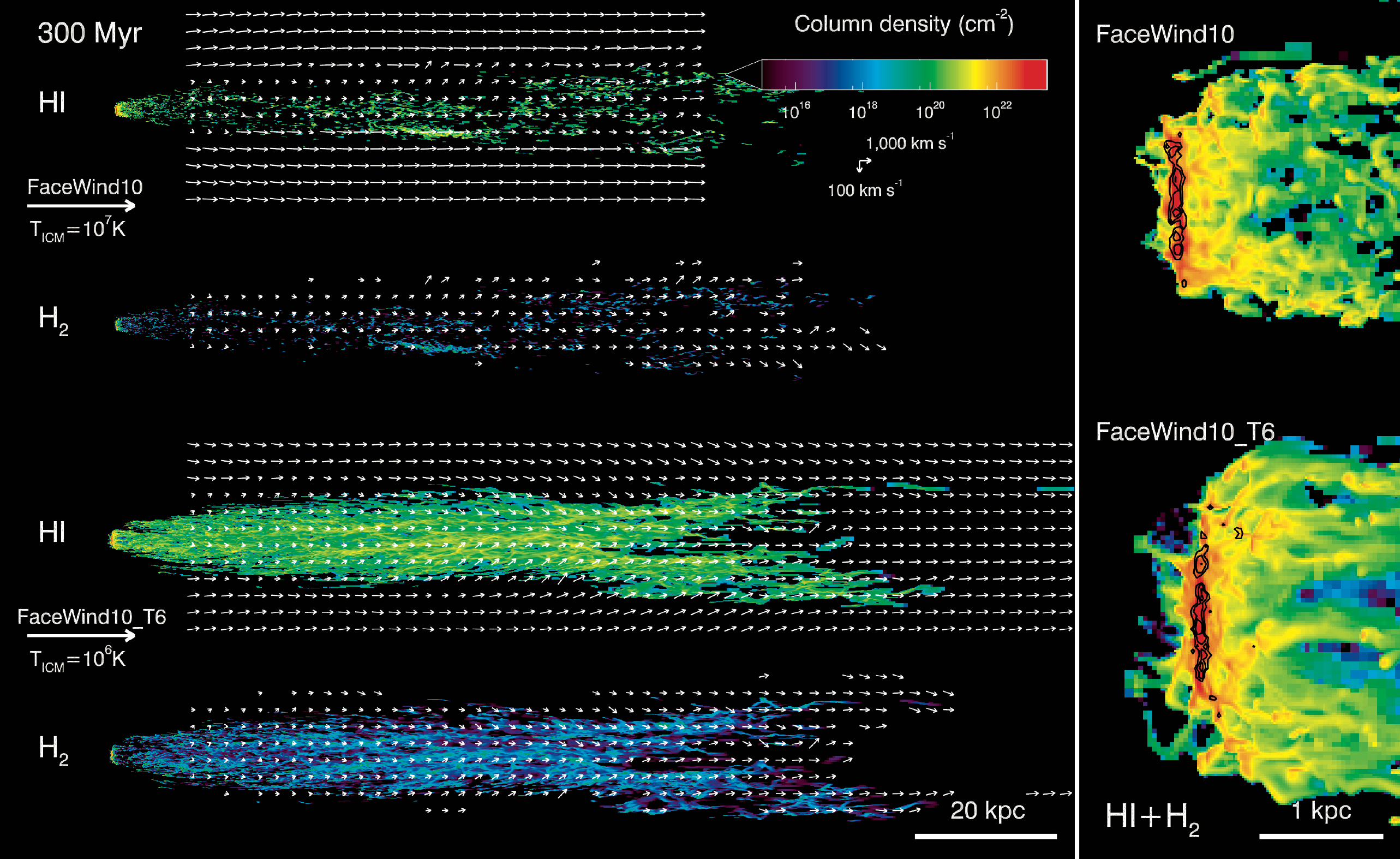}
\caption{Column densities and velocity fields of HI and H$_2$ in the \texttt{FaceWind10} (upper left panels, $T_{\rm ICM}=10^7\, {\rm K}$) and \texttt{FaceWind10\_T6} (bottom left panels, $T_{\rm ICM}=10^6 \, {\rm K}$) runs at 300 Myr. The velocity fields are shown only in the tail region if either the column density of HI or H$_2$ is greater than 0. The length of the arrow corresponds to the magnitude of the velocity, as indicated in the top right area. A significant amount of HI or H$_2$ gas is collapsed from the ICM when the cooling rates are increased by lowering the temperature of the ICM (see the text). Also included in the rightmost panels are zoomed images of the column density distribution of neutral hydrogen (HI+H$_2$). The black contours show the distribution of stars younger than 20 Myr. Even under such extreme conditions, no stars form in the massive tail but only in the disk plane, suggesting that extraplanar star formation may require even more gas-rich and/or massive galaxies in clusters.} 
\label{fig:img_z_strong}
\end{figure*}

\subsection{Lack of star formation in the stripped tails}
\label{sec:strong_tail}

 \citet{kapferer09} investigated the star formation of a simulated galaxy ($M_*=2\times10^{10}\,\msun$, $f_{\rm gas}=0.34$) experiencing an ICM wind with varying densities ($6\times10^{-5} \la n_{\rm H}\la 3\times10^{-3}\, {\rm cm^{-3}}$) and velocities ($100 < v_{\rm ICM} < 1000\,{\rm km\,s^{-1}}$), and found that the amount of newly formed stars for 500 Myr is increased by an order of magnitude due to ram pressure, compared to a galaxy experiencing no external ram pressure. In particular, the fraction of stars formed in the stripped tail is well correlated with the wind velocity. Among all the cases, the galaxy forms eight times more stars in a run where the ICM density and velocity are similar to our \texttt{FaceWind10} run, and more than 95\% of the young stars are found in the stripped wake. Therefore, the star formation is remarkably triggered in the tail, while the strong ram pressure suppresses the star formation in the disk.
\citet{tonnesen10} found a hint of star formation in a ram-pressure-stripped tail of a galaxy with $M_*=10^{11}\,\msun$ and $f_{\rm gas}=0.1$ hit by a moderate ICM wind of $P_{\rm ram}/k_{\rm B}\sim 6.4\times10^{4}\,{\rm K\,cm^{-3}}$, which is slightly stronger than the ram pressure of our moderate winds. They argue that the thermal pressure of the ICM mainly regulates the star formation in the wake of stripped galaxies once gaseous clouds are stripped from galaxies. The results of these experiments commonly suggest that the conditions of the ICM may be a key parameter controlling star formation activities in the stripped wakes.

In contrast, no significant star formation occurs in the wakes of our simulated galaxies regardless of the strength of the ram pressure. As mentioned in Section 3.1, the tail has more than $1.5\times10^6$ cells with the highest refinement level ($\Delta x=18$ pc) at $\sim$ 300 Myr in its stripped tail of \texttt{FaceWind10} run. For comparison, the numbers of the highest level cells are $\sim2.9\times10^5$ and $\sim7.4\times10^5$ in the tails of \texttt{FaceWind} and \texttt{EdgeWind} runs at the same epoch, respectively. Even though the strong ICM wind drives the formation of significant large-scale HI and H$_2$ tails, only a few new stellar particles are observed in the tail of  \texttt{FaceWind10}. We attribute the different conclusions to the different amounts of gas and/or different modelling strategies of star formation. The stellar mass of the simulated galaxies in \citet{kapferer09} and \citet{tonnesen10} are 10 and 50 times more massive than our galaxy ($M_*=2\times10^{9}\,\msun$), respectively, and their gas mass is also 4 and 6 times larger than that of our galaxy ($M_{\rm gas}=1.75\times10^{9}\,\msun$) as well, respectively. Therefore, a larger amount of gas may be stripped and provide the source for extraplanar star formation. We note, however, that the amount of newly formed stars is unlikely to be as significant as in \citet{kapferer09}, given that even low-density gas ($\nH < 1\,\cmq$) is allowed to convert into stars in their study. As mentioned earlier, we do find gas clumps as dense as $\nH\sim10\,\cmq$, but they are not thermodynamically unstable enough to collapse to form stars in our simulation. A more reliable conclusion can be reached only by simulating more massive galaxies in realistic environments with physically well motivated star formation and stellar feedback models, which we plan to carry out in the near future.


One may wonder whether the lack of star formation in the tail is due to inefficient cooling. Although our choice for the density and temperature of the ICM wind is motivated by observations and cosmological simulations, our set-up is admittedly idealized, and does not account for the possible interaction with colder and denser gas in  cool-core clusters \citep[e.g.,][]{hudson10}. To better understand the emergence of extraplanar star formation, we run an additional simulation by lowering the temperature of the ICM wind to $10^6\,{\rm K}$ (from the fiducial $10^7\,$K), which is equivalent to increasing the cooling rates by a factor of $\sim3$ in the ICM \citep[e.g.,][]{sutherland93}. Note that once the cooling time becomes much shorter than the free-fall time, the gas collapse will increase the density, hence the actual cooling rates can increase non-linearly. 

Interestingly, the amount of HI tail gas in the \texttt{FaceWind10\_T6} run is increased by an order of magnitude ($M_{\rm HI}=1.5\times10^9\,\msun$) at $t=300\,{\rm Myr}$, compared with the result from \texttt{Facewind10} ($M_{\rm HI}=1.9\times10^8\,\msun$), as illustrated in Figure~\ref{fig:img_z_strong}. During the initial phase of the stripping, the typical column density of HI in the vicinity of the galaxy is high ($N_{\rm HI}\sim10^{19}$--$10^{21}\,{\rm cm^{-2}}$). At this stage, mixing with the ICM barely occurs, and the tail is primarily composed of the stripped ISM. Once the stripped gas mixes with the hot ICM  by turbulent motions induced by the collision between the ISM and ICM, the typical temperature and density of the mixed gas ($f_{\rm HII}>0.5$ and $f_{\rm mix}=0.25$--$0.75$) becomes $T\sim10^{4}$--$10^{7}\,{\rm K}$ with median $T\sim4\times10^6$ K and $\nH \sim 10^{-3}$--$10^{-1}\,\cmq$. 
The cooling timescale of such gas is short \citep[$t_{\rm cool}\la 0.1$--$1\,{\rm Myr}$,][]{sutherland93} enough for $\sim40\%$ of the mixed clouds to overcome the cloud crushing \citep{armillotta16,armillotta17,gronke18}, which occurs approximately on a time scale of 0.1--1\,{\rm Myr}\footnote{The cloud crushing timescale is estimated as $\left(n_{\rm H,ISM}/n_{\rm H,ICM}\right)^{1/2} \, r_{\rm cloud}/v_{\rm ICM}$ \citep{gronke18}, where the typical size of the cloud ($r_{\rm cloud}$) is assumed to be on the order 100 pc.}.  As a result, a large amount of ICM HII turns into HI. For comparison, the mixed clouds in the tail in \texttt{FaceWind10} have $T\sim10^{4}$--$10^{7}\,{\rm K}$ with median $T\sim10^7$ K and $\nH \sim 10^{-2.5}$--$10^{-0.5}\,\cmq$, meaning $t_{\rm cool}\ga10 \,{\rm Myr}$ for $\sim 90\%$ of the mixed clouds. The huge difference in the cooling time scales of the mixed clouds results in the very different HI tail mass between the two runs. This process begins right behind the galaxy (see the right panels in Figure~\ref{fig:img_z_strong}), although the majority of the newly cooled HI appears to form at $ \ga 20 \, {\rm kpc}$. However, the exact location of the massive tail is likely to depend on the cooling timescale of the stripped ISM, and a larger amount of HI might form even at shorter distances if the galaxy were more gas-rich.

The column density of HI in the thick tails from the \texttt{FaceWind10\_T6} run often exceeds the typical transition density from HI to H$_2$ \citep[$N_{\rm HI}\sim 10^{20-21}\,{\rm cm^{-2}}$,][]{tumlinson02,wolfire08,lee15}. Furthermore, because the level of the Lyman-Werner flux from the galaxy is low and it is thus difficult to photo-dissociate molecular hydrogen \citep[e.g.,][]{gnedin09,sternberg14}, not only HI but also a significant quantity of molecular hydrogen is formed in the tail. Up to $\sim 10^8\,\msun$ of molecular hydrogen is observed in the tail in \texttt{FaceWind10\_T6}, whereas only a trace amount of molecular hydrogen ($\sim 10^7\,\msun$) is carried by the strong face-on wind with $T=10^7\,{\rm K}$. However, still no stars are formed in the extended HI and H$_2$ structures. The typical gas surface density at $\sim 40\, {\rm kpc}$ away from the galaxy center is $\Sigma_{\rm gas}\sim 1$--$30 \,{\rm \msun\, \rm pc^{-2}}$  or $N_{\rm H}\sim 10^{20}$--$3\times10^{21} \,{\rm cm^{-2}}$, and the average hydrogen number density is only about $\nH\sim1\,{\rm cm^{-3}}$.  Some gas in the tail reaches up to $\nH \sim 100\,\cmq$, but we find that the turbulent gas flow with densities $\nH\ge10\,\cmq$ is supersonic ($\mathcal\sim3$ and the virial parameter is very high ($\alpha_{\rm vir}\sim30$). The predicted star formation efficieny per free-fall time in these relatively dense tail is $\epsilon_{\rm ff}\sim10^{-6}$.  Consequently, it is difficult for star formation to occur  in the tail structure of a small galaxy, and reproducing jellyfish-like features \citep[e.g.,][]{poggianti17,sheen17} may require interactions between the ICM and the ISM of a more massive and gas-rich galaxy.

It is also worth noting that the tails of the two ram pressure stripped galaxies, D100 in Coma cluster (Abell 1656) and EOS137-001 in Norma cluster (Abell 3627), are found to be H$_2$-rich but HI deficient \citep{jachym17,jachym19}. These molecular tails are estimated to be massive ($\sim10^9\,M_{\odot}$) and extend to $\sim60$ kpc from the galaxy, but the level of star formation appears to be rather insignificant ($\sim10^{-3}-10^{-2}\,{\rm M_{\odot}\,yr^{-1}}$). This is qualitatively consistent with our findings from the \texttt{FaceWind10\_T6} run where the presence of the large amount of molecular clouds in the tail does not guarantee vigorous star formation. More interestingly, \citet{jachym17,jachym19} claim that the molecular tails are likely formed {\it in-situ}, rather than directly stripped from the galaxies because the age of the tails is estimated to be higher ($\gtrapprox200\,{\rm Myr}$) than the typical lifetime of dense molecular clouds ($\sim1-10\,$Myr). We do find that H$_2$ can form {\it in-situ}, but this is accompanied by an even more pronounced structure of HI, posing a challenge to our theoretical understanding of ram pressure stripping. Given that dust features are clearly seen in the disk of both galaxies and the inner tail of EOS137-001, the formation of H$_2$ may be enhanced due to dust, but our simulations fail to reproduce such H$_2$-rich but HI-deficient tails. 
Additional studies will be needed to pin down physical mechanisms behind the formation of H$_2$-rich but HI-deficient ram pressure stripped tails.


\begin{figure}
\centering 
\includegraphics[width=8.5cm]{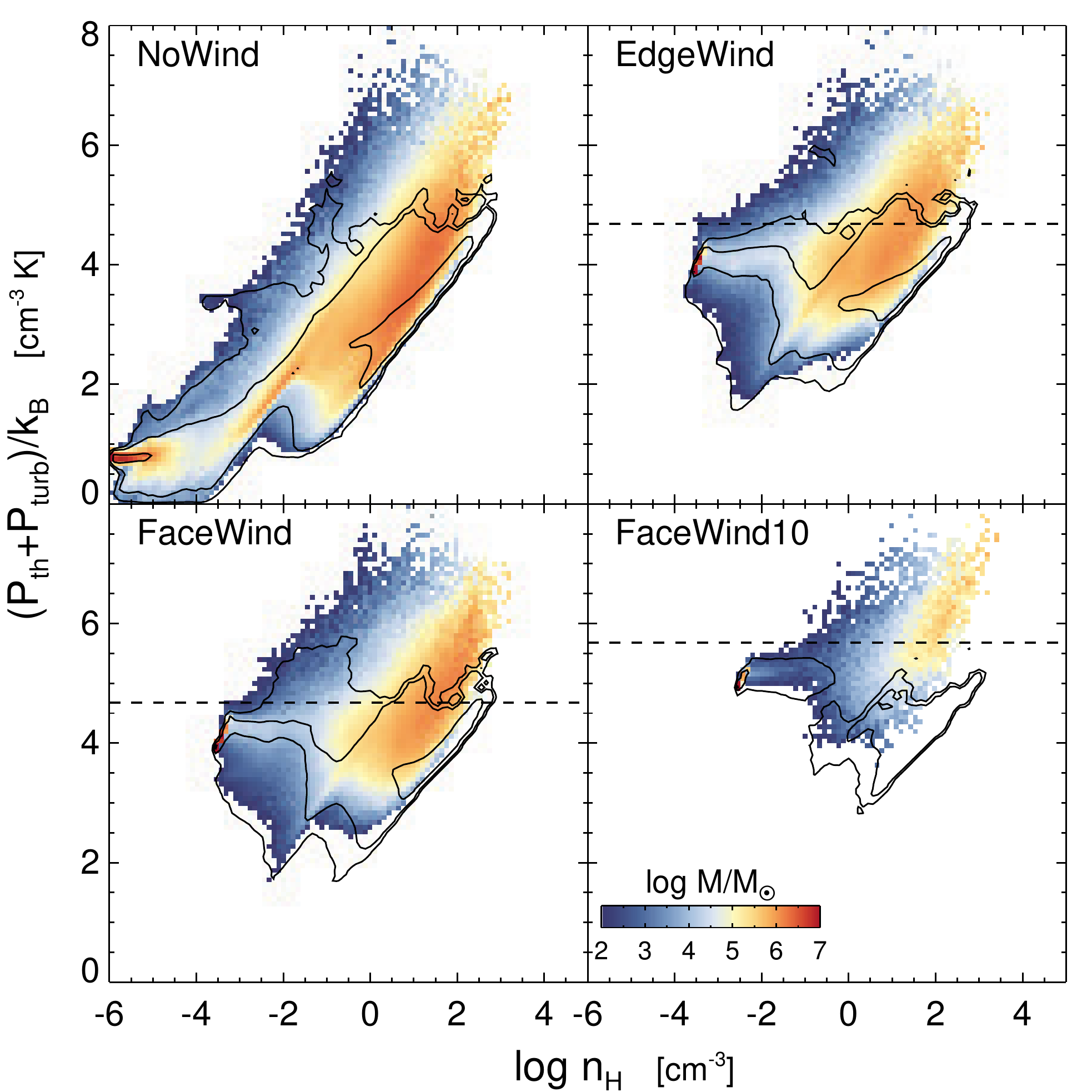}
\caption{Comparison between the total (thermal plus turbulent) pressure and the ram pressure of the ICM wind (dashed lines) at $t=600\, {\rm Myr}$.  Only the cells that are located in the front side of the wind are included in this plot. Turbulent pressure is computed as $P_{\rm turb}=\rho_{\rm gas} \sigma_{\rm gas}^2$, where $\sigma_{\rm gas}$ is the local gas velocity dispersion, as defined in Section 2.1.1. Different colors indicate the fraction of mass in each bin, as indicated in the legend. Black contours mark the thermal pressure (i.e. without turbulent pressure). It can be seen that ram pressure dominates over thermal pressure, but extra pressure provided by turbulence can counter-balance gas stripping due to the ICM wind.} 
\label{fig:img_phase_turb}
\end{figure}

\begin{figure}
\centering 
\includegraphics[width=8.5cm]{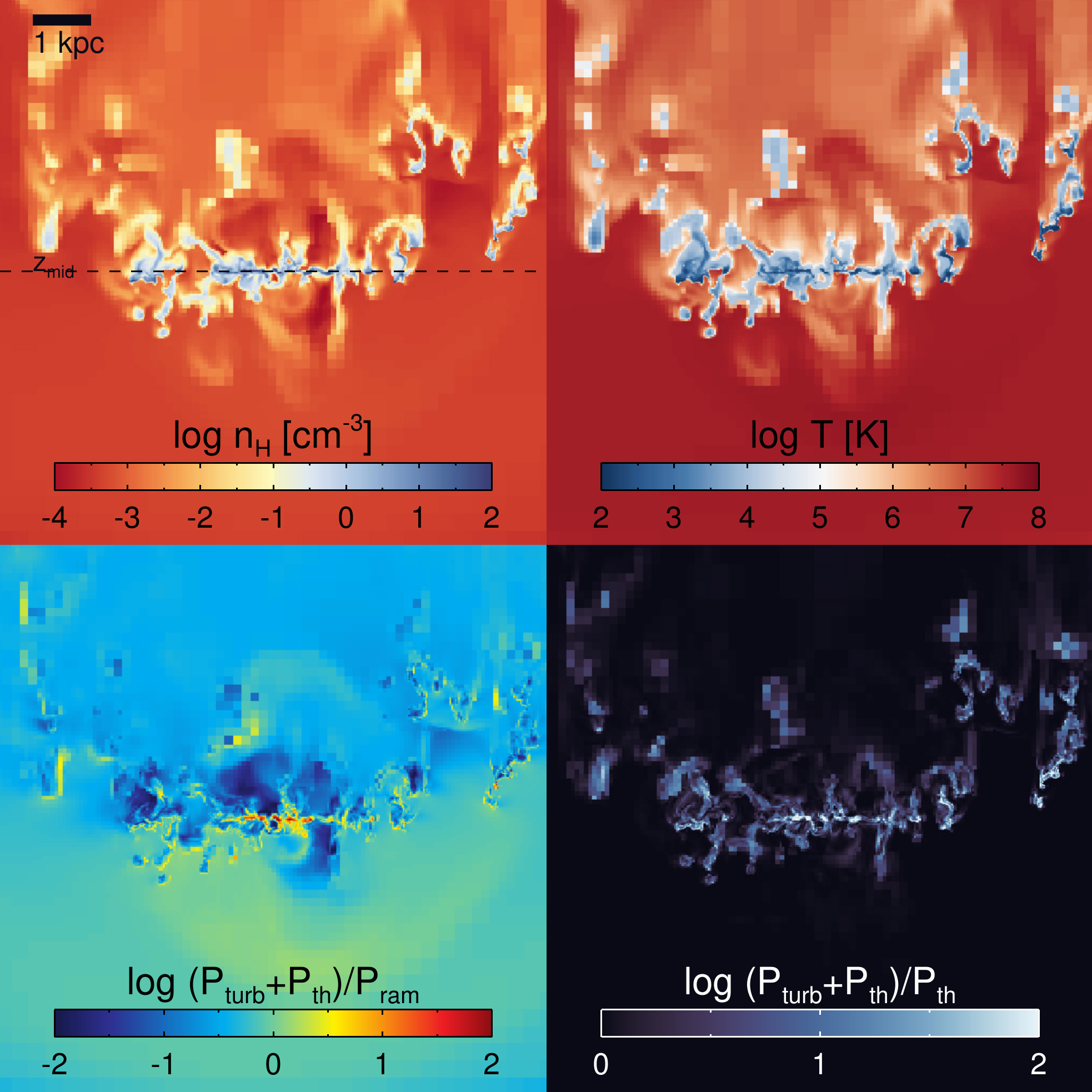}
\caption{Density (top left) and temperature (top right) distribution of a slice from the \texttt{FaceWind} run at $t=600\,{\rm Myr}$. The mid-plane of the simulated disk is marked as a dashed line. The bottom left panel indicates the relative strength of the total pressure (turbulent+thermal) and  ram pressure. The relative strength of the total pressure and thermal pressure is shown in the bottom right panel. Gas components with the total pressure less than the ram pressure are pushed away from the disk, while the dense, central disk with significant turbulent support is little affected by the ICM wind. }
\label{fig:img_slice}
\end{figure}

\subsection{Effects of turbulent ISM on ram pressure stripping}
\label{sec:pturb}

The ram pressure we apply ($P_{\rm ram}/k_{\rm B}=5\times10^4$ or $5\times10^5 \,{\rm K \,cm^{-3}}$) is higher than the typical thermal pressure of the ISM \citep[e.g.,][]{ostriker10}.  This naturally raises a question of why the ISM located at the front side of the ICM wind in the \texttt{FaceWind} or \texttt{FaceWind10} run is not completely stripped (e.g., Figure~\ref{fig:sfr_face}), especially given that the direction of gravitational restoring force is aligned with ram pressure? 

To understand this, we plot the total (thermal+turbulent) pressure of the gas in the runs with different ICM wind at $t=600\,{\rm Myr}$ in  Figure~\ref{fig:img_phase_turb}.  Note that we include only the cells with $0<x<x_{\rm mid}$ in case of \texttt{EdgeWind} or $0<z<z_{\rm mid}$ for other runs, i.e. cells that are located in the front side of the wind, so that the gas pressure is the only term that acts against ram pressure. Here $x_{\rm mid}$ or $z_{\rm mid}$ is the disk center of mass along x or z direction, determined iteratively within 3 kpc in radius. The turbulent pressure is computed as $P_{\rm turb}=\rho \, \sigma_{\rm turb}^2$, where the local turbulent velocity ($\sigma_{\rm turb}$) is measured as is done to estimate the star formation efficiency in Section 2.1.1. The ram pressure of the wind is included as black dashed lines. This figure demonstrates that the majority of the thin black contours, indicating the thermal pressure of the ISM, is indeed placed below the ram pressure of the moderate or strong wind (shown as the dashed lines in each panel). However, the moderate ICM wind barely penetrates the ISM and increases the contribution to the total column density for the central 100 pc slab by more than 10\% in Figure~\ref{fig:gas_face} (shown as white contours). This strongly suggests that there is additional pressure acting against the ram pressure. 

We argue that turbulent motions set by stellar feedback are crucial to the understanding of gas stripping in cluster environments. As can be seen in Figure~\ref{fig:img_phase_turb}, the turbulent pressure is greater than thermal pressure by more than an order of magnitude, especially at  $\nH\ga 10 \,\cmq$. This is not surprising, given that the motion of molecular clouds is known to be highly supersonic \citep[e.g.,][]{roman-duval10}. 
Although our resolution is not sufficient to resolve the detailed structure of the turbulent ISM, our simulated galaxy is still mainly supported by turbulence, which can counter-balance ram pressure in the front side of the wind before the gas is pushed away from the mid-plane of the disk and gravitational restoring force operates. To illustrate this, we present the distribution of density, temperature, and relative strength of turbulent pressure for the central slice of the \texttt{FaceWind} galaxy at $t=600\,{\rm Myr}$ in Figure~\ref{fig:img_slice}. The figure shows that the central dense disk supported by turbulent pressure is virtually unaffected by the wind, whereas the gas with lower total pressure is gradually stripped. For example, one can see that relatively dense ($\nH\sim1\,\cmq$) gas clumps located in the rightmost area are just being stripped, as their turbulent support is not strong enough to act against ram pressure\footnote{In the lower left panel of Figure~\ref{fig:img_slice}, the total pressure at the interface between the ICM wind and the ISM sometimes exceed the ram pressure, but these features are likely to be an artifact of the way we measure the velocity dispersion, as velocities change rapidly.}. In contrast, there also exists ISM  components with $\nH\la 1\,\cmq$ at $z<z_{\rm mid}$ which are not yet stripped despite their lower total internal pressure than the ram pressure. Visual inspection confirms that these gas clouds are relatively newly formed due to stellar feedback and stripped within $\sim 50 \, {\rm Myr}$. 


If turbulent pressure plays a vital role in delaying gas stripping, how can the simple estimate without an explicit dependence on turbulence by \citet{gunn72} give a reasonable estimate for the actual stripping radius, as shown in Section 3.1.1? This may be attributed to the self-regulated star formation in the simulated galaxies. Studies have shown that the mid-plane pressure, which is the sum of the thermal and turbulent pressure, is set by the heating rate, i.e. star formation rate, which is again driven by gravitational pull determined by the surface density \citep[e.g.,][]{blitz06,ostriker10}. This implies that galaxies with higher surface density have stronger pressure, which can counterbalance the stronger ram pressure, explaining the trends we observe in our simulations. We defer the precise determination of the effects of turbulence on ram pressure, again because our simulated ISM are unlikely to capture detailed turbulence structures, but we should be able to address this issue using higher-resolution simulations by controlling the level of turbulence in the future.

\subsection{Caveats and limitations}

Our results based on idealized wind tunnel simulations are qualitatively consistent with observational studies in the sense that HI gas is stripped more efficiently than H$_2$, and that star formation in the disk is possibly enhanced for a few hundred million years once the ICM wind hits the galaxy. However, as \citet{tonnesen19} noted, the impact of ram-pressure stripping in real clusters where the wind strength varies with time may be less dramatic than the continuous wind case. For example, galaxies that approach the center of a cluster from the outskirts will be affected by gradually increasing ram pressure, and most cluster satellites are likely to lose a significant fraction of their initial gas mass by the time they reach the central region of the cluster. Moreover, pre-processing in galaxy groups may lower the gas fraction even before a galaxy falls into the cluster \citep{jung18}, suggesting that cluster satellites may exhibit less notable stripped features in the central region of a cluster than presented in the \texttt{FaceWind10} run. Conversely, strong tidal fields, not considered in our simulations, disturb gas and stellar kinematics \citep[e.g.,][]{moore96}, possibly amplifying the gas loss, along with ram pressure. The precise determination of the two combined effects would require simulations of galaxies orbiting around a live dark matter halo filled with hot gas.

Another missing physical ingredient in this study is magnetic fields. The ICM as well as spiral galaxies are known to be magnetized at up to a few $\mu G$ level at the cluster center \citep{govoni04,vogt05,bonafede10,kuchar11}, with the field aligned along the spiral arm in the latter case \citep{beck05}. Motivated by observational results, several studies carried out magneto-hydrodynamic simulations to understand the role of magnetic fields in galaxies experiencing ram pressure stripping. Although physical conditions and simulation set-ups vary between simulations, it has been suggested widely that magnetic fields only weakly suppress the stripping rates and smooth the stripped features in the tail \citep{ruszkowski14,shin14,tonnesen14,ramos-martinez18}. This implies that magnetic fields can suppress the formation of clumpy clouds in ram-pressure-stripped tails, potentially making the star formation in the tail more difficult to reproduce.

Finally, resolution could be a factor influencing our results. Although we employ a high spatial resolution ($\Delta x_{\rm min}=18\,{\rm pc}$) with aggressive refinement criteria, it may not be sufficient to resolve hydrodynamic instabilities. In this case, gas stripping could be enhanced further \citep[e.g.,][]{agertz07}, although it is unclear how instabilities under ram pressure react to turbulence and magnetic tension \citep[e.g.,][]{ryu00}. Because of the high computational costs of the radiation-hydrodynamic simulations, we cannot perform convergence test at a higher resolution elements in the present study; however, we plan to investigate the issue by refining only a small volume of the tail in future work.

\section{Conclusions}
Motivated by observational studies that reveal different distributions of neutral and molecular gas in cluster satellites \citep[e.g.,][]{lee17, jachym19}, we have carried out a set of idealized simulations of a multi-phase star-forming galaxy with the ICM environments, which mimic cluster outskirts (moderate) and central regions (strong). For this, we have used a slightly modified version of the publicly available radiation-hydrodynamics code, \ramsesrt, which can trace the formation, dissociation, and ionization of HI and H$_2$ in the presence of strong stellar feedback \citep{kimm17,katz17}. We have examined the effects of the ICM wind under varying pressure conditions to investigate if star formation is triggered or quenched under different cluster environments. Our results can be summarized as follows. 

\begin {enumerate}

\item Moderate ICM winds can strip not only a significant amount of HI but also H$_2$ gas in the outer part of the simulated galaxies. The mass loss mainly occurs in the diffuse and intermediate-density ISM ($\nH\la10\,\cmq$), most of which becomes ionized as it is displaced from the galaxy (40 kpc, Figure~\ref{fig:outflow_rate}). 


\item Out of the two runs with the moderate ICM wind, the decrease in gas mass and size is more significant in the face-on wind case than in the edge-on wind (Figure~\ref{fig:mass_size}). In total, 39~\% and 15~\% of cold gas is stripped in the moderate face-on and edge-on wind runs, respectively, over a 500-Myr period. Conversely, the strong face-on wind removes more than 90\% of cold gas from the galaxy within 200 Myr.

\item Star formation is enhanced (40--80\%) in the presence of moderate winds, compared to the isolated case (Figure~\ref{fig:sfr}). In particular, the edge-on wind strongly boosts star formation at the interface between the wind and the ISM (Figure~\ref{fig:sfr_face}). In contrast, the severe mass loss due to the strong face-on wind quenches star formation, except in the central region ($r_c<0.3\,{\rm kpc}$, Figure~\ref{fig:csfr_evo}).

\item Gas clouds as dense as $\sim 10\,\msun \,{\rm pc^{-2}}$ are ubiquitous on the shielded side of the galaxy; However, little star formation occurs in the stripped tails in all runs. Even when a lower ICM temperature ($T_{\rm ICM}=10^6$ K) is applied to enhance radiative cooling in the ram-pressure-stripped tail, a negligible amount of stellar particles is formed, even though the amount of cold gas is increased by an order of magnitude (Figure~\ref{fig:img_z_strong}).

\item We argue that turbulent pressure in the multi-phase ISM provides support against ram pressure.  Nevertheless, the truncated radii of the gaseous disk in all the runs with the face-on ICM winds are found to be consistent with the simple size estimate based on the balance between the gravitational restoring force and the ram pressure \citep{gunn72}, as turbulent support is likely set by gravity through star formation and associated feedback.

\end {enumerate}

Our results are consistent with previous observational studies that found ram pressure stripped neutral hydrogen and molecular clouds traced by CO emission lines \citep[e.g.,][]{chung09,jachym19}, a hint of shrinking molecular disks \citep{lee17}, and an enhancement of star formation \citep[e.g.,][]{vulcani18} in cluster satellites. However, as discussed above, our simulations are highly idealized in terms of environments, and do not cover a wide range of mass that facilitates comparison with previous simulations  \citep {kapferer09,tonnesen10}. Furthermore, star-forming signatures from stripped wakes that characterize jellyfish galaxies \citep[e.g.,][]{sheen17} are not reproduced in the present study. Such outstanding issues call for future studies that can model the ICM-ISM for a massive, gas-rich galaxy in realistic clusters.


\section*{acknowledgments}
We thank Ivy Wong, Aeree Chung, and Yonghwi Kim for useful discussion. We also thank the anonymous referee for constructive comments and careful reading of the manuscript. TK was supported in part by the National Research Foundation of Korea (NRF-2017R1A5A1070354 and NRF-2020R1C1C100707911) and in part by the Yonsei University Future-leading Research Initiative (RMS2-2019-22-0216), and acted as the corresponding author. 
The supercomputing time for numerical simulations was kindly provided by KISTI (KSC-2017-C2-0039), and large data transfer was supported by KREONET, which is managed and operated by KISTI. This work was also performed using the DiRAC Data Intensive service at Leicester, operated by the University of Leicester IT Services, which forms part of the STFC DiRAC HPC Facility (www.dirac.ac.uk). The equipment was funded by BEIS capital funding via STFC capital grants ST/K000373/1 and ST/R002363/1 and STFC DiRAC Operations grant ST/R001014/1. DiRAC is part of the National e-Infrastructure.

\appendix
\section{Choice of H$_2$ clumping factor}

\begin{figure}
\centering 
\includegraphics[width=8.5cm]{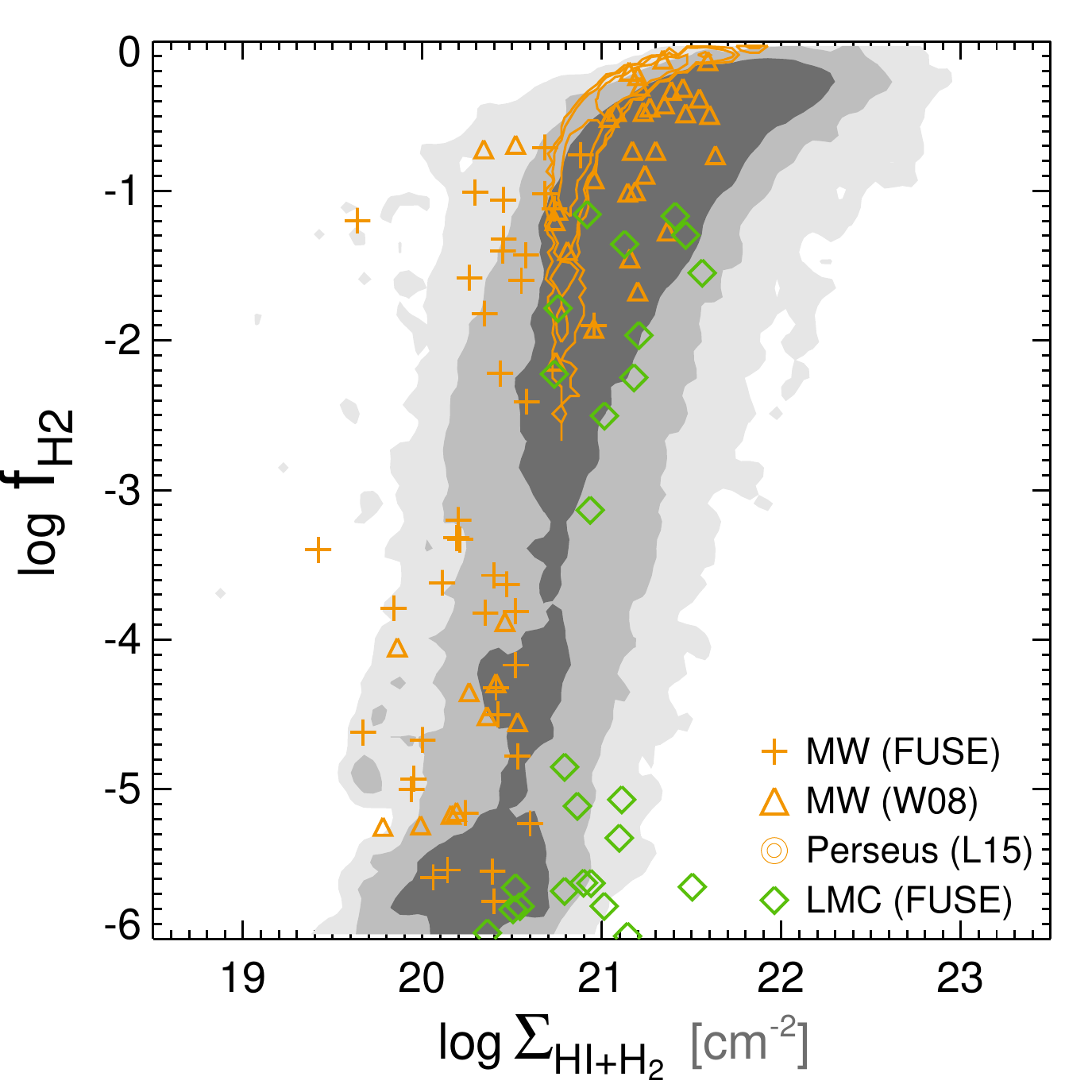}
\caption{Transition between neutral and molecular hydrogen as a function of the total hydrogen column density. The results for the Milky Way (MW) halo obtained from the Far Ultraviolet Spectroscopic Explorer (FUSE) data are shown as the orange cross symbol \citep{browning03}, while the orange triangles display the results for the sight-lines along the Galactic plane \citep{wolfire08}. Also included as the orange contours are the Perseus molecular cloud measurements by \citet{lee15}. The FUSE results for the Large Magellanic Cloud (LMC) are presented as green diamonds \citep{tumlinson02}. Our results from the \texttt{NoWind} run with $Z=0.75\,{\rm Z_{\odot}}$, shown as gray shaded contours (1, 2, and 3 $\sigma$), are consistent with the average MW and LMC trends.  }
\label{fig:transition}
\end{figure}

In this work, we adopt a clumping factor of $C_{\rho}=3$ to compute the formation rate of molecular hydrogen, motivated by \citet{gnedin09}. This is a lower value than the original clumping factor used by \citet[][$C_{\rho}=10$]{gnedin09}, because our computational resolution is slightly better than that of the previous work. Figure~\ref{fig:transition} indeed shows that the galaxies simulated with  $Z=0.75\,{\rm Z_{\odot}}$ reasonably match the mean trends measured from the Milky Way halo \citep{browning03}, Milky Way disk \citep{wolfire08}, Perseus molecular cloud \citep{lee15}, and Large Magellanic cloud \citep{tumlinson02}. 

\begin{figure*}[h]
\centering 
\includegraphics[width=0.9\textwidth]{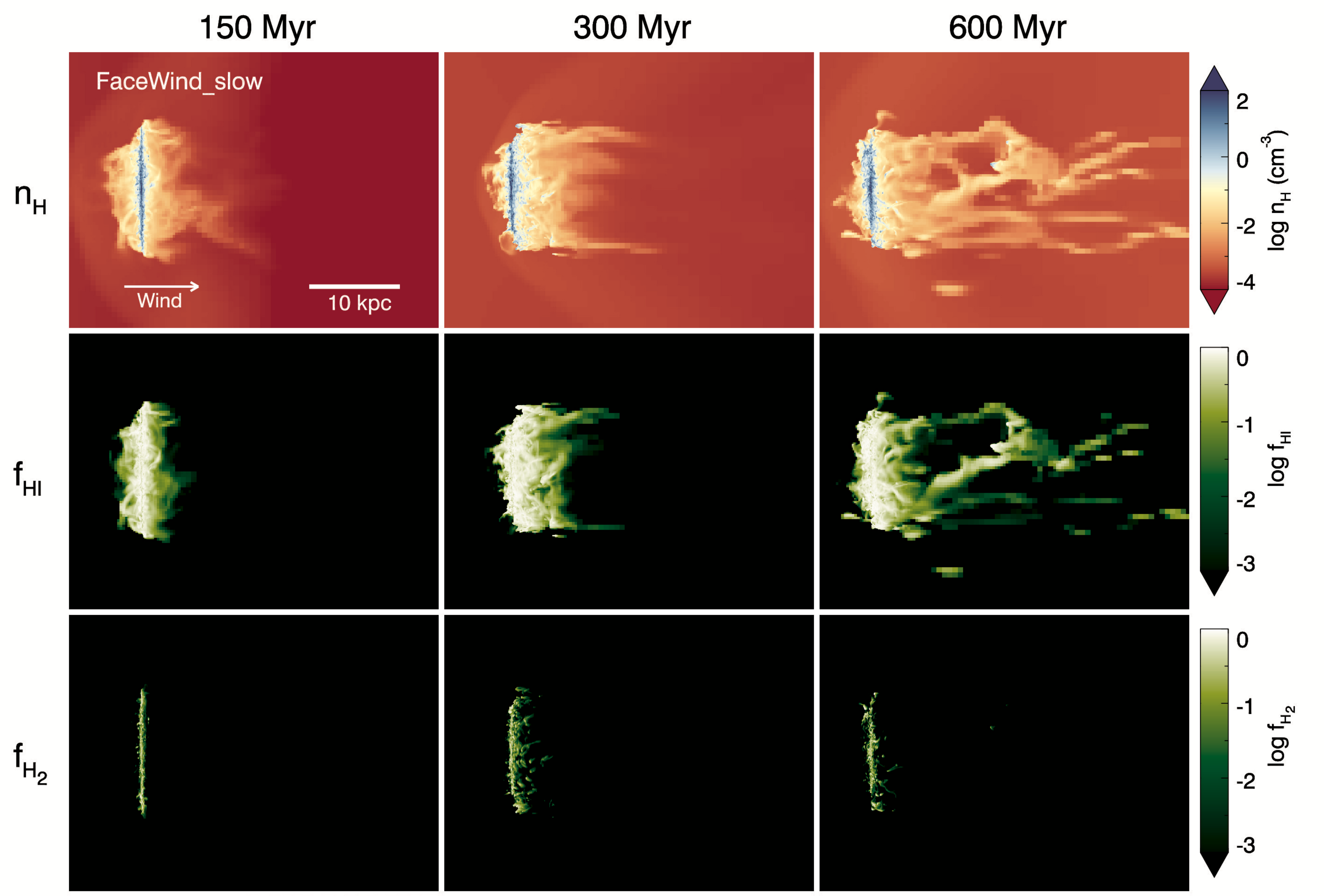}
\caption{ Same as Figure~\ref{fig:img1}, but for the run with the slow ($v_{\rm ICM}=300\,{\rm km\,s^{-1}}$) face-on wind (\texttt{FaceWind\_slow}). The weak ram pressure mildly strips the CGM, forming HI tails that are much smoother than those of the moderate or strong face-on wind runs (see Figures~\ref{fig:img1}--\ref{fig:img2}). H$_2$ is barely stripped, and only perturbed by the slow ICM wind.}
\label{fig:img_slow}
\end{figure*}

\section{Impact of weak ram pressure stripping}

It is known that galaxies experience ram pressure from far outside the virial radius of a cluster \citep[e.g.,][]{jung18}. We thus test the effects of very weak RAM pressure with $P_{\rm ram}/k_{\rm B}=4\times10^3\,{\rm cm^{-3}\, K}$ by adopting a slower ICM velocity of 300 \,\kms\ and $\nH=3\times10^{-4}\,\cmq$ (\texttt{FaceWind\_slow}). The temperature of the ICM wind in \texttt{FaceWind\_slow} is lowered in order to suppress the acceleration due to its thermal pressure. Note that the ram pressure of the slow ICM wind corresponds to $\approx 10\%$ of $P_{\rm ram}$ applied in the \texttt{FaceWind} run.

Figure~\ref{fig:img_slow} shows the projected gas distributions of a galaxy stripped with the slow wind. Because the ram pressure of the slow ICM wind is weak, the stripping takes longer and the stripped tails appear less dramatic than in other runs with moderate or strong winds. Instead, they show rather a smooth HI distribution near the galaxy, with no extraplanar star formation. 

\begin{figure}
\centering 
\includegraphics[width=10.5cm]{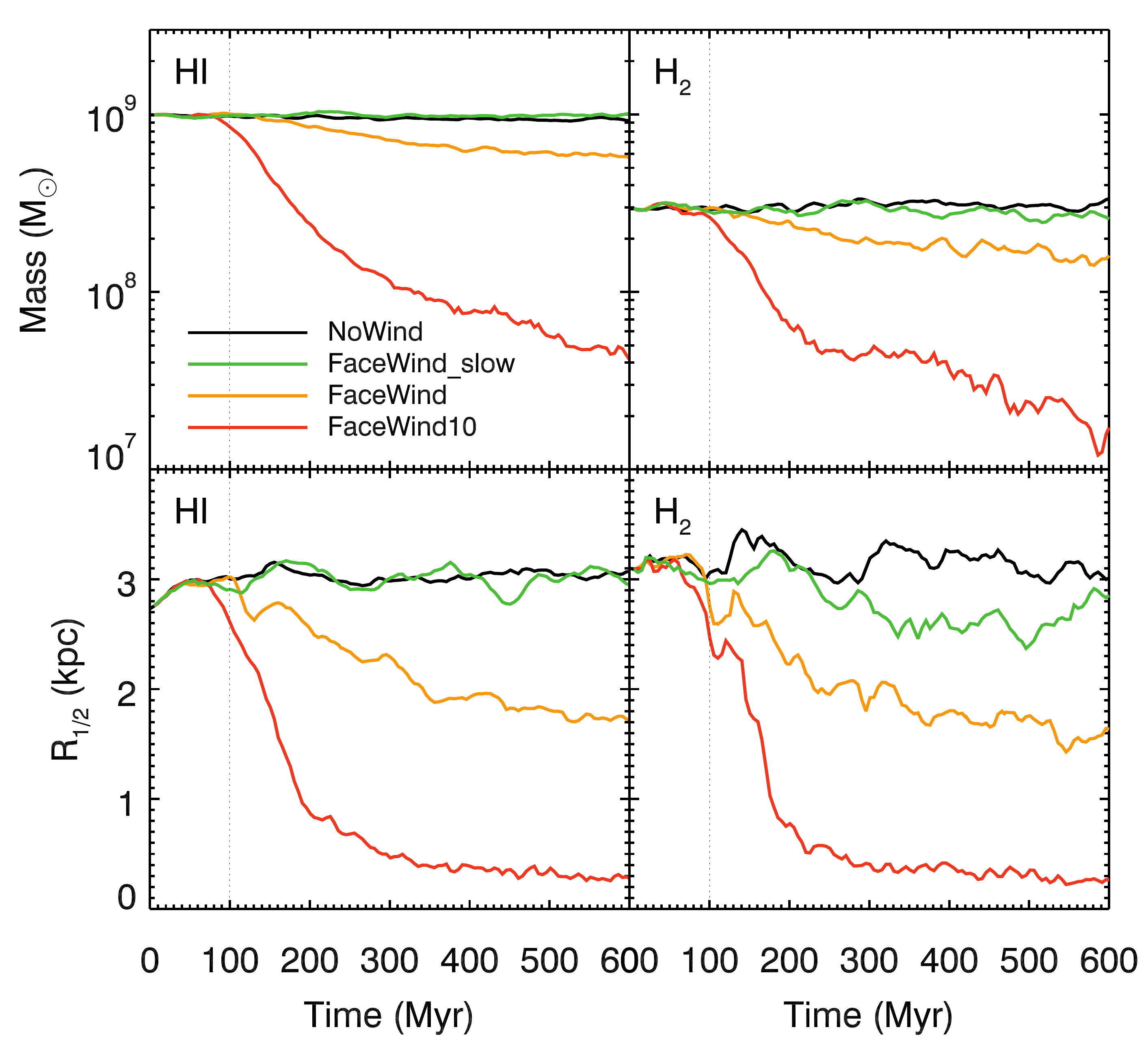}
\includegraphics[width=7cm]{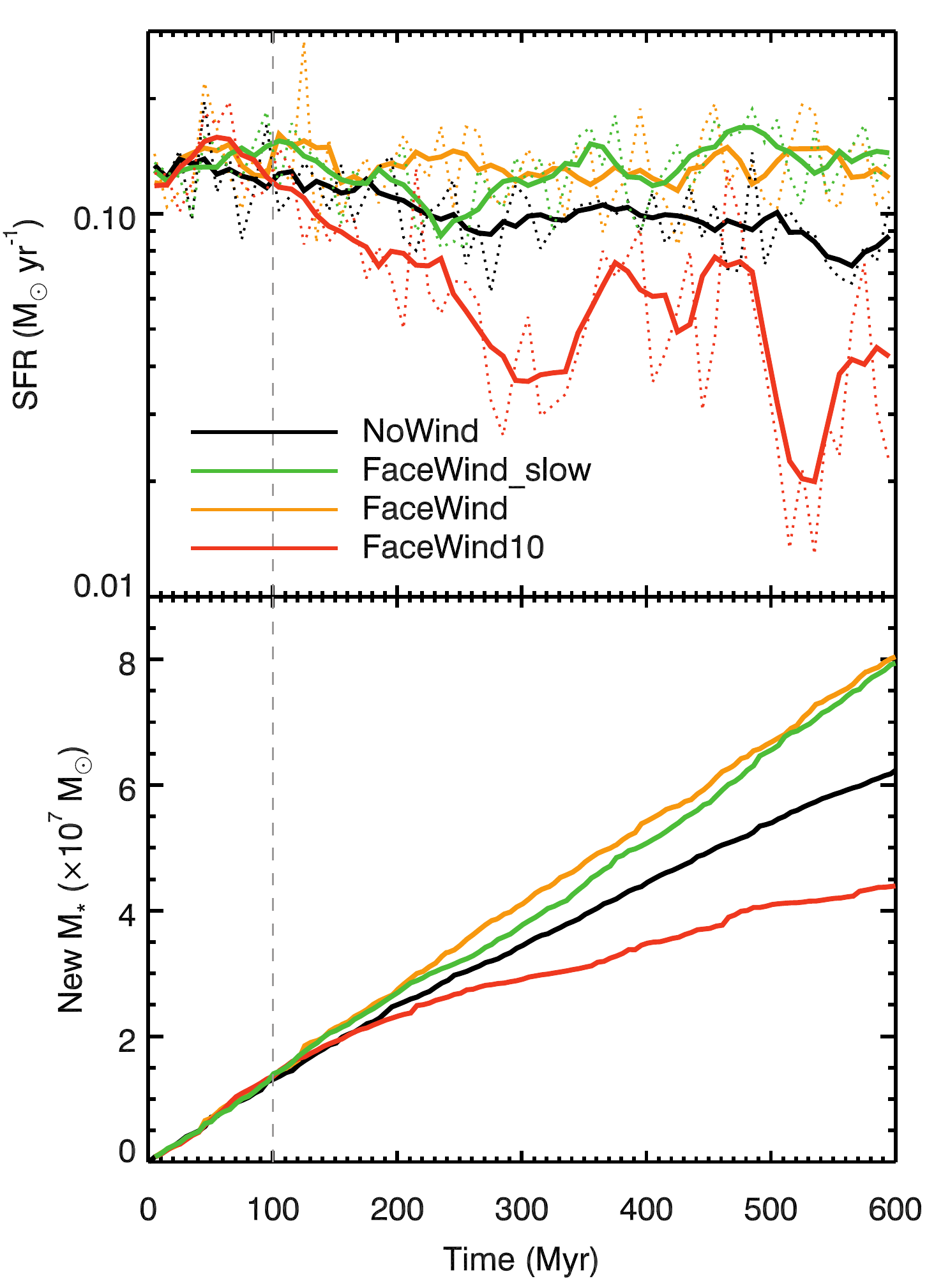}
\caption{{\it Left:} the evolution of the gas mass in HI and H$_2$ within the cylindrical volume of radius $r=10\,{\rm kpc}$ and height $|z|= 3 \, {\rm kpc}$ centered on the galaxy (top panels). The evolution of the half-mass radii ($R_{1/2}$) of the HI and H$_2$ disks is also shown in the bottom panels. In the \texttt{FaceWind\_slow} run, HI and H$_2$ disks appear to be little affected by the slow ICM wind. {\it Right:} same as Figure~\ref{fig:sfr}, but with three different face-on wind runs and isolated case, for comparison. The presence of the slow ICM wind enhances the star formation similarly to the \texttt{FaceWind} run.}
\label{fig:prop_slow}
\end{figure}

It seems that the weak ram pressure insignificantly affects the HI and H$_2$ disks, compared with those in the \texttt{FaceWind} and \texttt{FaceWind10} runs (Figure~\ref{fig:prop_slow}, left panels). When the galaxy is first hit by the slow ICM wind, the CGM is removed gradually, while the dense ISM stays rather intact. But once the CGM is stripped, the ram pressure gently compresses the gas in the ISM, and the star formation becomes enhanced after $\sim 200\,{\rm Myr}$.  Eventually, the galaxy under the weak ICM wind forms significantly more stars than the isolated case (Figure~\ref{fig:prop_slow}, right panels), as is the case for the \texttt{FaceWind} run. Our experiment thus suggests that it is possible to observe galaxies with enhanced star formation at the periphery of their host dark matter halo. However, the exact level of enhancement in star formation should be determined in a more realistic setting where the properties of the ICM winds vary along the orbital motion of the ram pressure stripped galaxy \citep{tonnesen19}.

\end{document}